\documentclass[aps,superscriptaddress,onecolumn,amsmath,amssymb]{revtex4}
\usepackage{preamble} 

\begin{document}

\title{Shaping nonlinear optical response using nonlocal forward Brillouin interactions}

\author{Shai Gertler}
\email[]{shai.gertler@yale.edu}
\affiliation{Department of Applied Physics, Yale University, New Haven, CT 06520, USA}

\author{Prashanta Kharel}
\affiliation{Department of Applied Physics, Yale University, New Haven, CT 06520, USA}

\author{Eric A. Kittlaus}
\affiliation{Department of Applied Physics, Yale University, New Haven, CT 06520, USA}
\affiliation{Jet Propulsion Laboratory, California Institute of Technology, Pasadena, CA 91109, USA}

\author{Nils T. Otterstrom}
\affiliation{Department of Applied Physics, Yale University, New Haven, CT 06520, USA}

\author{Peter T. Rakich}
\email[]{peter.rakich@yale.edu}
\affiliation{Department of Applied Physics, Yale University, New Haven, CT 06520, USA}


\begin{abstract}
We grow accustomed to the notion that optical susceptibilities can be treated as a local property of a medium. 
In the context of nonlinear optics, both Kerr and Raman processes are considered local, meaning that optical fields at one location do not produce a nonlinear response at distinct locations in space. 
This is because the electronic and phononic disturbances produced within the material are confined to a region that is smaller than an optical wavelength. 
By comparison, Brillouin interactions can result in a highly nonlocal nonlinear response, as the elastic waves generated through the Brillouin process can occupy a region in space much larger than an optical wavelength.  
The nonlocality of these interactions can be exploited to engineer new types of processes, where highly delocalized phonon modes serve as an engineerable channel that mediates scattering processes between light waves propagating in distinct optical waveguides.
These types of nonlocal optomechanical responses have been recently demonstrated as the basis for information transduction, however the nontrivial dynamics of such systems has yet to be explored. 
In this work, we show that the third-order nonlinear process resulting from spatially extended Brillouin-active phonon modes involves mixing products from spatially separated, optically decoupled waveguides, yielding a nonlocal `joint-susceptibility'.
We further explore the coupling of multiple acoustic modes and show that multi-mode acoustic interference enables a tailorable nonlocal-nonlinear susceptibility, exhibiting a multi-pole frequency response.

\end{abstract}

\maketitle
\section{\label{sec:intro}Introduction}
Optical nonlinear processes such as four-wave-mixing and harmonic generation are usually described in terms of a frequency dependent, spatially local susceptibility. In other words, the optical fields in one location do not alter the nonlinear response in another point in space \cite{boyd_book}. In Raman scattering, the short mean-free path of the THz-frequency optical phonons participating in the nonlinear process also results in a local susceptibility \cite{boyd_book}. By comparison, nonlocal nonlinearities require a mechanism by which the optical fields in one location affect the optical response in another location. Nonlocal response has been studied in the context of thermally induced effects \cite{dabby1968thermal,horowitz1992large,rotschild2006long}, as well as more exotic systems such as nematic liquid crystals \cite{izdebskaya2018stable}, trapped atoms \cite{shahmoon2016highly}, Rydberg gases \cite{sevinccli2011nonlocal}, plasmonic systems \cite{pollard2009optical,krasavin2016nonlocality}, and graphene \cite{fakhri2015nonlocal}. All of these interactions are the result of transport mechanisms, such as heat, electric charge, or atoms, that mediate the optical response over an extended distance.

Alternatively, phonons that participate in a Brillouin scattering process can serve as the transport mechanism for long-range interactions. These acoustic modes can be long-lived and propagate multiple optical wavelengths before decaying, yielding nonlocal dynamics \cite{renninger2018bulk,shin2013tailorable,kittlaus2016_FSBS}. This acousto-optic process is a three-wave mixing process producing a coherent interaction of optical waves and acoustic phonons \cite{boyd_book,wiederhecker2019brillouin,eggleton2019brillouin}. More specifically, in a forward-Brillouin scattering process the optical fields are co-propagating, while the phonons produced by the scattering process are emitted perpendicular to the direction of optical wave propagation \cite{shin2013tailorable,van2015interaction,kittlaus2016_FSBS}. The transverse nature of the phonons, combined with their long lifetime, enables them to explore a space much larger than the acousto-optic overlap region \cite{shin2013tailorable}. This allows the design of structures where the phonons interact with multiple optical fields, which are otherwise optically decoupled \cite{shin2015control,kittlaus2018rf}.

In this work, we analyze a theoretical model of these forms of nonlocal interactions in the context of Brillouin-active superstructures supporting acoustic modes extending between multiple optical waveguides. Because the phonon mode taking part in the Brillouin scattering process has an overlap with multiple optical fields, the scattering processes in the different waveguides are no longer independent. Hence, the light scattering in one waveguide will affect the dynamics in another, resulting in a `joint-susceptibility' between the two spatially separated waveguides. We show that in the case of negligible optical dispersion, the intensity beat-note of light propagating through a waveguide in the device results in pure phase modulation of a spatially separate optical guided wave. This phase modulation is determined by the acoustic and optical properties in both waveguides, revealing the nonlocal nature of the interaction. We further extend our analysis to the case of multiple acoustic modes taking part in the acousto-optic process. Specifically, we show that coupling multiple acoustic modes results in phonon super-modes, all occupying the extended space and interacting with the optical fields. The coherent interference of these phonon super-modes yields a multi-pole frequency response for the joint-susceptibility, showing a faster frequency roll-off compared to a typical acoustic Lorentzian lineshape.

Such nonlocal Brillouin interactions have been recently demonstrated both in optical fiber and in chip-scale photonic devices. In multi-core fibers, light guided within spatially distinct cores can be coupled to acoustic modes occupying the entire fiber cross section \cite{diamandi2017opto}. By comparison, integrated photonic systems allow additional structural degrees of freedom which can be used to tailor both the optical and acoustic modes that participate in the nonlocal interaction \cite{shin2015control,kittlaus2018rf}. These devices demonstrate new behaviors which can be very useful as filters \cite{kittlaus2018rf}, transducers \cite{shin2015control}, oscillators \cite{diamandi2018highly}, and modulators \cite{kittlaus2018non} for both optical and microwave applications.
\section{\label{sec:theory}Theoretical study}
\begin{figure}
    \centering
    \includegraphics[scale=0.8]{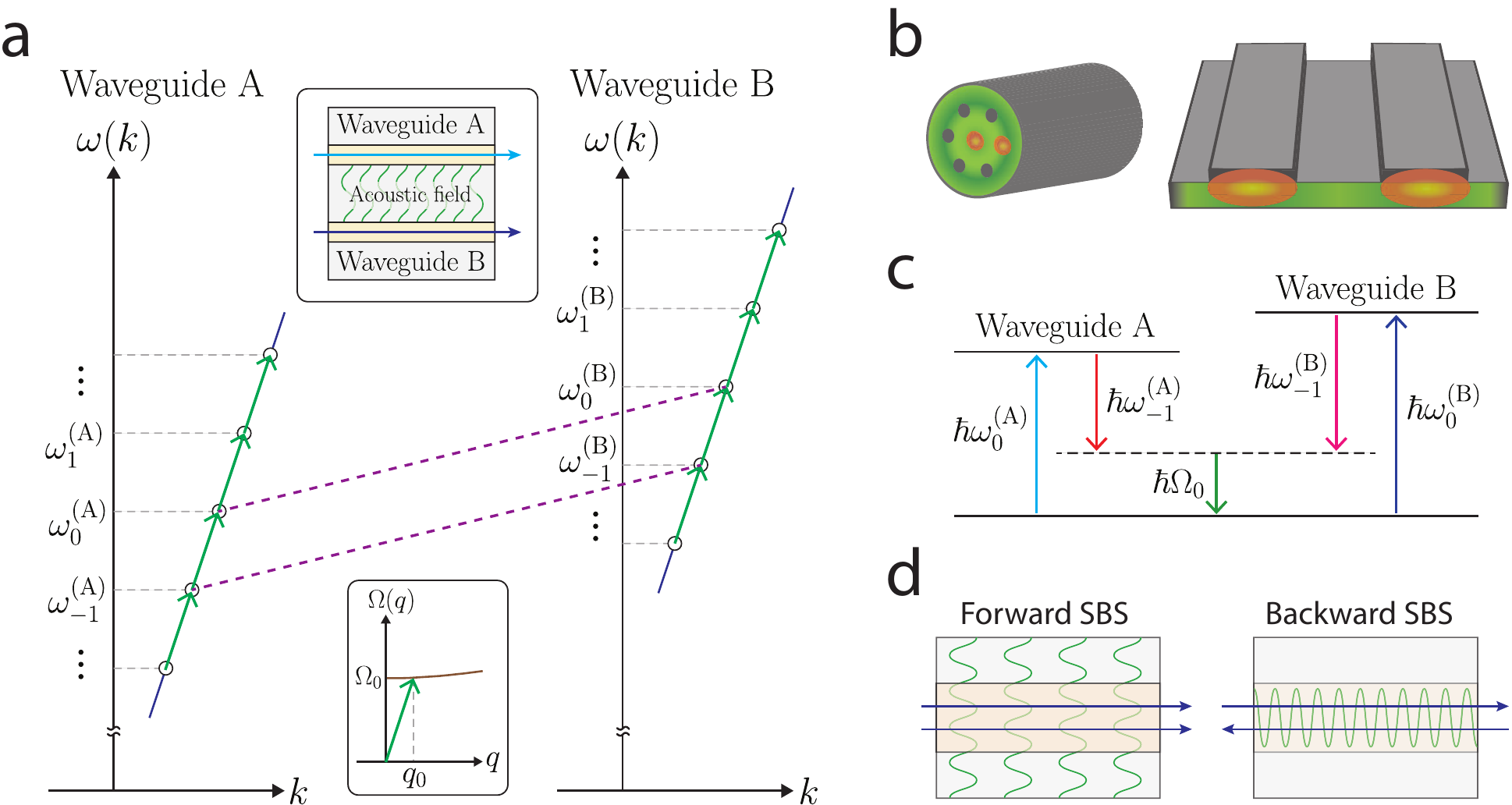}
    \caption{
    a. Dispersion diagram of an optical mode in each of the two waveguides. The phonon coupling between optical tones (green arrow) is phase matched in both waveguides. The phonon couples multiple tones spaced by frequency $\Omega_0$, as long as the dispersion relation is linear.
    Bottom inset: The acoustic dispersion diagram, illustrating the acoustic mode taking part in the interaction is near its cutoff frequency with a vanishing axial wave-vector. 
    Top inset: A schematic illustration of the described system, with two guided optical modes, both coupled to an extended acoustic mode.
    b. Left: A multi-core optical fiber, where optical modes are spatially separated (orange), and a transverse acoustic mode (green) is supported by the entire cladding cross-section, such as described in Ref. \cite{diamandi2017opto}. Right: An illustration of a chip-scale device with two optical waveguides each supporting an optical mode (orange), and a single spatially-extended acoustic mode (green), discussed in Ref. \cite{kittlaus2018rf,kittlaus2018non}. 
    c. Energy level diagram illustrating two optical tones in each waveguide interacting with a spatially extended nonlocal acoustic mode.
    d. Schematic comparison between forward and backward Brillouin processes. In a forward geometry, the transverse nature of the acoustic modes enable them to explore a large space, extending well beyond the acousto-optic overlap. In a backward process utilizing a bulk acoustic mode, the acoustic and optical waves are confined to a similar space.
    }
    \label{fig:intro_fig}
\end{figure}

We begin our analysis by considering a system consisting of two optical waveguides and supporting a single acoustic mode overlapping with both optical waveguides. While the two waveguides are optically decoupled, the light in each waveguide is coupled to the acoustic mode through a forward stimulated-Brillouin scattering (FSBS) process. 
The optical fields are co-propagating, and the photon-phonon coupling is a consequence of electrostrictive forces and radiation pressure \cite{rakich2012giant,sipe2016hamiltonian}, which can be tailored through the design of the device geometry \cite{rakich2010tailoring,shin2013tailorable}.
Examples of such system are lithographically defined chip-scale devices, as well as multi-core optical fiber, illustrated in Fig. \ref{fig:intro_fig}(b). 

The FSBS process in each of the waveguides can be described as a three-wave interaction between two photon and a phonon, as illustrated in the phase-matching diagrams in Fig. \ref{fig:intro_fig}(a).
The phase matching condition required in both waveguides is $q(\Omega_0)=k(\omega_0)-k(\omega_{-1})$, where $k(\omega)$ is the optical wave-vector at frequency $\omega$, and $q(\Omega_0)$ the acoustic wave-vector at frequency $\Omega_0$. Since the optical waves have similar wave-vectors, this requires a cut-off phonon mode, with a vanishing axial wave-vector, such that the acoustic wave is perpendicular to the direction of the optical wave propagation \cite{kang2009tightly,qiu2013stimulated,kharel2016_Hamiltonian}.
The transverse nature of these phonons allows them to extend much further than the optical waveguide. This is in contrast with backward-Brillouin scattering processes utilizing bulk acoustic modes, where the phonon occupies a similar region as the optical waves, illustrated in Fig. \ref{fig:intro_fig}(d). The wavelengths of the light in the two waveguides can be different, as long as energy is conserved in the process, illustrated in Fig. \ref{fig:intro_fig}(c). This requires $\hbar\Omega_0 = \hbar\omega_0^\text{(A)} - \hbar\omega_{-1}^\text{(A)}$ and $\hbar\Omega_0 = \hbar\omega_0^\text{(B)} - \hbar\omega_{-1}^\text{(B)}$, such that the phonon frequency matches the frequency difference of the two optical tones in both waveguides A and B. Considering both the phase matching and energy conservation, this requires the optical modes in both waveguides to have the same optical group velocity.

In the absence of strong optical dispersion, an FSBS process enables light to be cascaded to multiple optical frequencies \cite{kang2009tightly,kharel2016_Hamiltonian,kittlaus2016_FSBS,wolff2017cascaded}, and we describe the optical fields in each waveguide as a sum of discrete tones, spaced by a frequency $\Omega$. The equations of motion describing this system can be derived using quantum operators following  \cite{sipe2016hamiltonian,kharel2016_Hamiltonian} and are outlined in Appendix \ref{subsec:app_Hamiltonian}. Alternatively, a classical approach, where classical field amplitudes are calculated using a non-linear polarization wave equation is derived in Appendix \ref{subsec:app_Classical}. 

We denote the steady-state optical field amplitudes in the two waveguides as \({a}^\text{(A)}_n\) and \({a}^\text{(B)}_n\) with frequencies \(\omega^\text{(A)}_n\) and \(\omega^\text{(B)}_n\). We assume a constant group velocity over the frequency range of interest, such that the optical tones can cascade to an arbitrary number of sidebands. We denote the steady-state acoustic field amplitude as \(b\), and the acoustic dissipation rate \(\Gamma\). The equations of motion for the optical fields in each waveguide and the acoustic field are given by
\begin{equation}
    \begin{split}
    \frac{\partial {a}^\text{(A)}_n}{\partial z} &= -\frac{i}{v^\text{(A)}}\left( {g^\text{(A)}} b {a}^\text{(A)}_{n-1} + {g^\text{(A)}}^* b^{\dagger}{a}^\text{(A)}_{n+1}\right) ,\\
    \frac{\partial {a}^\text{(B)}_n}{\partial z} &= -\frac{i}{v^\text{(B)}}\left( {g^\text{(B)}} b {a}^\text{(B)}_{n-1} + {g^\text{(B)}}^* b^{\dagger}{a}^\text{(B)}_{n+1}\right) ,\\
    b &= -i\left(\frac{1}{i\Delta + \Gamma/2}\right)\sum_{n} \left({g^\text{(A)}}^* {a^{\dagger}}^\text{(A)}_n {a}^\text{(A)}_{n+1} + {g^\text{(B)}}^* {a^{\dagger}}^\text{(B)}_n {a}^\text{(B)}_{n+1} \right),
    \end{split}
    \label{eq:a}
\end{equation}
where \(\Delta = \Omega_0-\Omega\) is the detuning of the modulation frequency from the acoustic resonance \(\Omega_0\), and we have neglected optical loss. The summation is implied to cover all sidebands generated in the process, \(v^\text{(A)}\) and \(v^\text{(B)}\) are the optical group velocities in each of the waveguides, and \(g^\text{(A)}\) and \(g^\text{(B)}\) are the acousto-optic coupling rates. The field amplitudes are normalized such that the propagating optical and acoustic power is given by $P_n = \hbar \omega_0 v a^\dagger_n a_n$ and $P_\text{ac} = \hbar \Omega_0 v_\text{ac} b^\dagger b$ respectively \cite{kharel2016_Hamiltonian}. More details of the derivation are presented in Appendix  \ref{subsec:app_Hamiltonian}. Under these conditions, the phonon field is constant in space, i.e. \(\ \partial_z b(z) = 0\), and can be evaluated at \(z=0\) yielding
\begin{equation}
    b = -i\left(\frac{1}{i\Delta + \Gamma/2}\right)\sum_{n} \left({g^\text{(A)}}^* {a^{\dagger}}^\text{(A)}_n {a}^\text{(A)}_{n+1} + {g^\text{(B)}}^* {a^{\dagger}}^\text{(B)}_n {a}^\text{(B)}_{n+1} \right) \biggr\rvert_{z=0} .
    \label{eq:phonon_field}
\end{equation}

Next, in order to study the dynamics of the system, we analyze the effect of light scattering in waveguide A on the spectral evolution of a separate optical wave propagating in waveguide B. To do this, we assume two tones in the input of waveguide A at frequencies \(\omega_0^\text{(A)}\) and \(\omega_{-1}^\text{(A)} = \omega_0^\text{(A)} - \Omega \), with amplitudes \(a_0^\text{(A)}\) and \(a_{-1}^\text{(A)}\), which will drive the acoustic mode through a forward-Brillouin interaction, as seen from Eq. (\ref{eq:phonon_field}). Assuming a single tone in the input of waveguide B, the second term of Eq. (\ref{eq:phonon_field}) does not contribute, and we are left with
\begin{equation}
    b = -i\left(\frac{1}{i\Delta+\Gamma/2}\right) {g^\text{(A)}}^* {a^\text{(A)}_{-1}}^{\dagger}(0) \  {a}^\text{(A)}_{0}(0).
    \label{eq:phonon_field_two_tones}
\end{equation}
Plugging into Eq. (\ref{eq:a}) gives the spatial evolution of ther optical fields
\begin{equation}
    \begin{split}
        \frac{\partial {a}_n^\text{(A)}}{\partial z}&= -\frac{1}{{v}^\text{(A)}} \left|g^\text{(A)}\right|^2 \left[ {a}_{n-1}^\text{(A)} \chi {a^\text{(A)}_{-1}}^{\dagger}(0) {a}^\text{(A)}_{0}(0) - {a}_{n+1}^\text{(A)} \chi^* {a}^\text{(A)}_{-1}(0) {a^\text{(A)}_{0}}^{\dagger}(0) \right],\\
        \frac{\partial {a}_n^\text{(B)}}{\partial z}&= -\frac{1}{{v}^\text{(B)}} \left[ g^\text{(B)} g^\text{(A)*} {a}_{n-1}^\text{(B)} \chi {a^\text{(A)}_{-1}}^{\dagger}(0) {a}^\text{(A)}_{0}(0) - g^\text{(B)*} g^\text{(A)} {a}_{n+1}^\text{(B)} \chi^* {a}^\text{(A)}_{-1}(0) {a^\text{(A)}_{0}}^{\dagger}(0) \right],
    \end{split}
\end{equation}
where we denote the frequency response as \( \chi = \left({i\Delta+\Gamma/2}\right)^{-1} \). 

In order to understand the nonlinear susceptibility produced by the forward-Brillouin process, we examine the $n=-1$ tone in each waveguide (the Stokes wave), and consider a small signal analysis, such that we can neglect tones with $|n|\geq2$, yielding
\begin{equation}
    \begin{split}
    \frac{\partial {a}_{-1}^\text{(A)}}{\partial z}&= i \gamma^\text{(A)} {a}_{0}^\text{(A)} {a}^\text{(A)}_{-1}(0) {a^\text{(A)}_{0}}^{\dagger}(0),    
    \end{split}
    \qquad
    \begin{split}
    \frac{\partial {a}_{-1}^\text{(B)}}{\partial z}&= i \gamma^\text{(B)} {a}_{0}^\text{(B)} {a}^\text{(A)}_{-1}(0) {a^\text{(A)}_{0}}^{\dagger}(0),
    \end{split}
\end{equation}
where we have defined the susceptibilities in the two waveguides
\begin{equation}
    \begin{split}
    &\gamma^\text{(A)} = -\frac{i}{{v}^\text{(A)}} \left|g^\text{(A)}\right|^2 \chi^*,
    \end{split}
    \qquad
    \begin{split}
    &\gamma^\text{(B)} = -\frac{i}{{v}^\text{(B)}} g^\text{(B)*} g^\text{(A)} \chi^*.
    \end{split}
\end{equation}
We see that this is a third order susceptibility, where we have three field amplitudes driving the field amplification, similar to the form derived for four-wave mixing and backward stimulated Brillouin scattering \cite{boyd_book}. Examining the equation for waveguide B reveals the nonlocal nature of the susceptibility, where field amplitudes in waveguide A determine the response in the optically decoupled, spatially separated waveguide B. Further, the susceptibility $\gamma^\text{(B)}$ depends on the Brillouin coupling rate in both waveguides A and B. Similar expressions can be written for the anti-Stokes ($n=1$) tone in the waveguides.

Next, we assume assume the two waveguides are similar, such that \(v^\text{(A)} = v^\text{(B)} = v\) and \(g^\text{(A)} = g^\text{(B)} = g\), valid in many physical systems \cite{shin2015control,kittlaus2018rf,kittlaus2018non}. The equations describing the fields in waveguides A and B are now
\begin{equation}
\begin{split}
    \frac{\partial {a}^\text{(A)}_{n}}{\partial z} =-\frac{1}{v} \left|g\right|^2 \left|{a}^\text{(A)}_{-1}(0) \  {a}^\text{(A)}_{0}(0)\right| \left|\chi\right| \left(a^\text{(A)}_{n-1} \ e^{i\phi}e^{i\Lambda} - a^\text{(A)}_{n+1} \ e^{-i\phi}e^{-i\Lambda}\right),\\
    \frac{\partial {a}^\text{(B)}_{n}}{\partial z} =-\frac{1}{v} \left|g\right|^2 \left|{a}^\text{(A)}_{-1}(0) \  {a}^\text{(A)}_{0}(0)\right| \left|\chi\right| \left(a^\text{(B)}_{n-1} \ e^{i\phi}e^{i\Lambda} - a^\text{(B)}_{n+1} \ e^{-i\phi}e^{-i\Lambda}\right),
\end{split}
\label{eq:reccurance_with_phases}
\end{equation}
where we denote the relative phase between the two tones at the input to waveguide A as \(\Lambda = \arg ({a^\text{(A)}_{-1}}^{\dagger}(0) \ {a}^\text{(A)}_{0}(0) ) \) and the phase of the frequency response \(\phi = \arg (\chi) \).
We now solve the differential equation by rotating the field operators \(\bar{a}_n = a_n e^{-i\left(\phi+\Lambda \right)n} \), giving us the form
\begin{equation}
    \frac{\partial \bar{a}_{n}}{\partial z} =-\frac{1}{2} G_\text{B} \sqrt{P_0^\text{(A)} P_{-1}^\text{(A)}} \ \frac{\Gamma}{2} \left|\chi\right| \left(\bar{a}_{n-1} - \bar{a}_{n+1} \right),
\end{equation}
for both waveguides, and we have expressed the input fields in terms of optical power, using \(P_n = \hbar\omega_n v a_n^{\dagger} a_n \), and the acousto-optic coupling in terms of Brillouin gain \(G_\text{B} = 4 |g|^2/(\hbar \omega v^2 \Gamma) \) \cite{kharel2016_Hamiltonian, wiederhecker2019brillouin}. 
This recurrence relation is consistent with the Bessel function identity \( J_n' = \frac{1}{2} ( J_{n-1} - J_{n+1} )\), such that the optical fields can be written as a linear combination \(\bar{a}_n(z) = \sum_m c_{n,m} J_m (- \xi \ z )\), where \( \xi  = G_\text{B} ( P_0^\text{(A)} P_{-1}^\text{(A)})^{1/2} (\Gamma/2) |\chi|\). We can find the coefficients \(c_{n,m}\) using \(J_m(0) = \delta_{m,0} \) such that \(c_{n,m} = \bar{a}_{n-m}(0) \), and using the identity \(J_n(-x) = J_{-n}(x)\), we arrive at 
%
%
%
\begin{equation}
        {a}_n(z) = \sum_m a_{n+m}(0) J_{m} \left(\xi \ z \right) e^{-i\left(\phi+\Lambda \right)m},
\end{equation}
where we have rotated the operators back to the field envelope frame. Plugging in the initial conditions for each waveguide gives us
\begin{equation}
    \begin{split}
        a_n^\text{(A)}(z) &=  a^\text{(A)}_{0}(0) J_{-n} \left(\xi \ z \right)e^{i\left(\phi+\Lambda \right)n} + a^\text{(A)}_{-1}(0) J_{-(n+1)} \left(\xi \ z \right) e^{i\left(\phi+\Lambda \right)(n+1)},\\
        a_n^\text{(B)}(z) &=  a_{0}^\text{(B)}(0) J_{-n} \left(\xi \ z \right)e^{i\left(\phi+\Lambda \right)n},
    \end{split}
    \label{eq:a4}
\end{equation}
where we have denoted the amplitude at the input to waveguide B $a_0^\text{(B)}$, with optical frequency $\omega_0^\text{(B)}$.
Evaluating the optical field amplitude at the output of waveguide B, i.e. the sum of the amplitudes at equally spaced frequencies \(s^\text{(B)}_\text{out}(t) \propto \sum_n {a}^\text{(B)}_n e^{-i (\omega^\text{(B)}_0 + n\Omega )t}\), we arrive at
\begin{equation}
    s^\text{(B)}_\text{out}(t) = \sqrt{P_{0}^\text{(B)}} e^{-i \omega^\text{(B)}_0 t}\sum_n J_{n} \left( \frac{\Gamma}{2} \left|\chi\right| G_\text{B} \sqrt{P_0^\text{(A)} P_{-1}^\text{(A)}} \ z \right) e^{-i \Big(\Omega t - \left( \phi + \Lambda\right) + \pi\Big)n},
    \label{eq:B_out_with_Lorentzian}
\end{equation}
where we have used the identity \(J_{-n} = (-1)^n J_n \), and neglected a global phase of the input field. The field amplitudes are normalized such that the total power entering waveguide B is $P_0^\text{(B)}$. Using the Jacobi-Anger expansion, \( \sum_n J_n(z) e^{in\varphi} = e^{iz\sin{\varphi}}\), we can directly see that the field is purely phase modulated
\begin{equation}
    s^\text{(B)}_\text{out}(t) = \sqrt{P_{0}^\text{(B)}} \ e^{-i\omega^\text{(B)}_0t} \ \exp \left[i  \frac{\Gamma}{2} \left|\chi\right| G_\text{B} \sqrt{P_0^\text{(A)} P_{-1}^\text{(A)}} \ z \sin{\Big(\Omega t - \left( \phi + \Lambda\right)\Big)} \right],
\end{equation}
where the modulation depth is determined by the
Brillouin gain, the optical powers in waveguide A and the propagation length. The frequency response $\chi(\Omega)$ of this phase-modulated field follows a Lorenzian lineshape, determined by the acoustic resonant frequency and dissipation rate, with a magnitude given by
\(|\chi|^2 = [(\Omega_0 -\Omega)^2+(\Gamma/2)^2]^{-1}\).

\begin{figure}
    \centering
    \includegraphics[scale=.8]{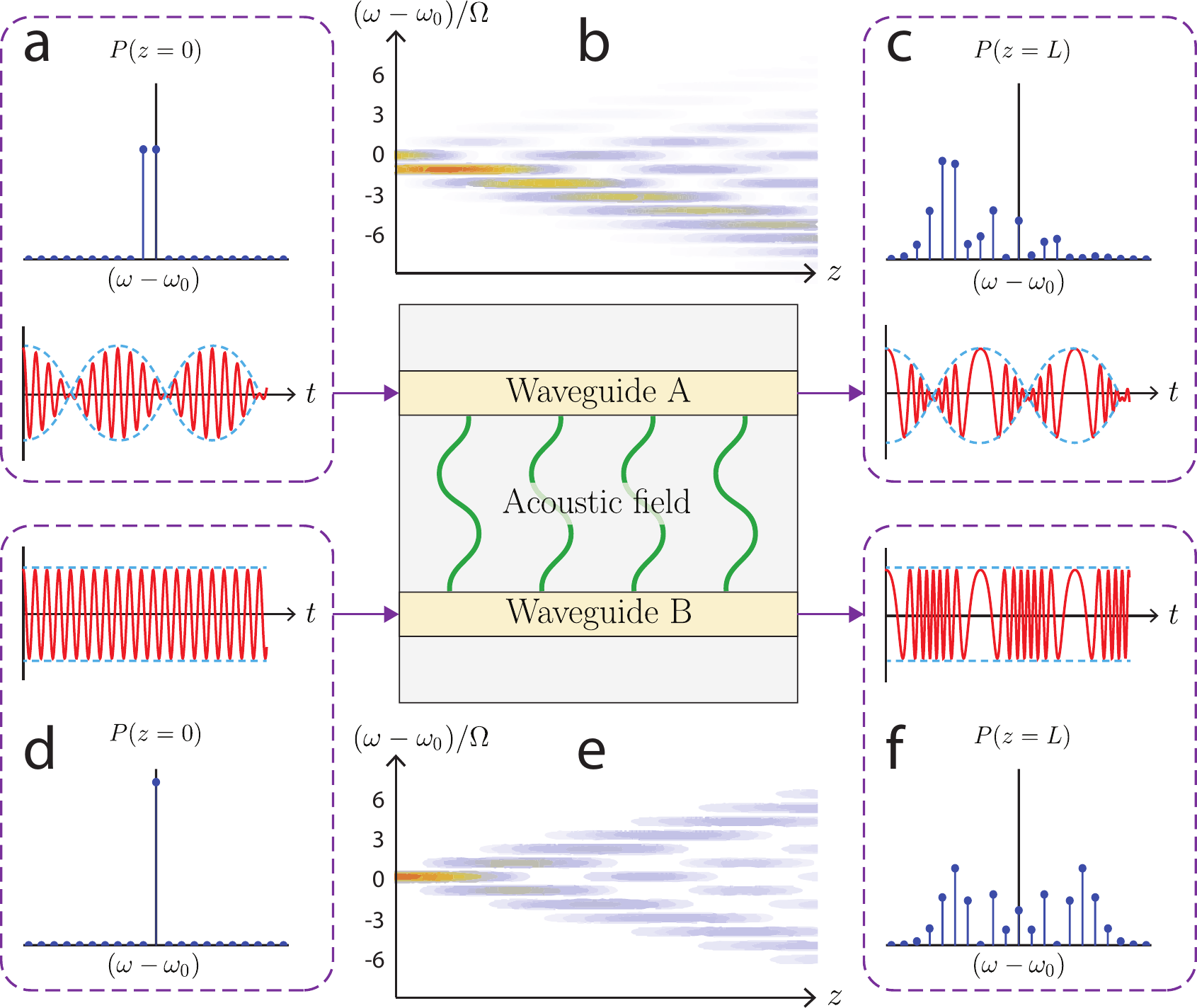}
    \caption{
    Simulation of the nonloncal transduction using a FSBS-active device: 
    a. The input to waveguide A is comprised of two tones, illustrated both in the frequency and time domain showing the resulting envelope modulation.
    b. The optical fields drive a phonon field, scattering light into multiple sidebands as they propagate along the device. c. The field at the output of waveguide A shows phase modulation of the input wave, however the intensity envelope is unaffected. The same steps are illustrated for waveguide B: 
    d. A single CW tone is launched, e. is scattered by the acoustic field as it propagates along the device, and 
    f. shows the phase modulation side-bands at the output.
    The time-domain illustrations in these figures are schematic, not shown to scale. 
    }
    \label{fig:two_tones}
\end{figure}
Summarizing the results of this section, we see how the optical intensity beat note generated by the two tones in waveguide A drive the acoustic mode, as seen in Eq. (\ref{eq:phonon_field}) and illustrated in Fig. \ref{fig:two_tones}(a). This acoustic field modulates an optical tone in waveguide B as it propagates through the device length, Fig. \ref{fig:two_tones}(b), resulting in pure phase modulation, Fig. \ref{fig:two_tones}(f).
The optical field in waveguide A also experiences phase modulation as it interacts with the acoustic field, as illustrated in Figs. \ref{fig:two_tones}(b) and \ref{fig:two_tones}(c), however the intensity beat note is not affected, discussed further in Appendix \ref{subsec:app_emit}. This is also consistent with the acoustic mode envelope being constant in space, as we saw in Eq. (\ref{eq:phonon_field}).
We can also notice in Figs. \ref{fig:two_tones}(b) and \ref{fig:two_tones}(c) that there is an overall red-shift of the light in waveguide A. This is a result of the finite lifetime of the phonons in the acoustic mode.In the absence of optical loss, the overall power dissipated is the energy dissipated from the acoustic field, given by $P_L=\Gamma \hbar \Omega b^\dagger b L $. This can also be expressed in terms of the optical powers and the Brillouin gain as $P_L = (\Omega/\omega_0^\text{(A)}) G_\text{B} P_0^\text{(A)} P_{-1}^\text{(A)} L$. 
Our treatment assumed no optical dispersion in the frequency range over which the optical tones are cascaded. As a result, we have used a single group velocity, and assumed an infinite summation in Eq. (\ref{eq:a}) and the following derivation. In the absence of this assumption, the output field will no longer be purely phase modulated, and will exhibit some amplitude modulation \cite{wolff2017cascaded}, as is discussed further in Appendix \ref{subsec:app_dispersion}. Another case where optical cascading is absent is when two different optical spatial modes take part in the nonlinear process \cite{kang2010all,kittlaus2017_Intermodal}, and was recently utilized to demonstrate a nonlocal response \cite{kittlaus2018non}. This type of inter-modal scattering is further discussed in Appendix \ref{subsec:app_XMPPER}.

\subsection*{\label{subsec:multi_membrane}Multiple coupled phonon modes}
The frequency response of the Brillouin-induced nonlinear susceptibility is determined by the properties of the phonon taking part in the interaction. 
This response can be further engineered by utilizing multiple acoustic modes, such that they all contribute to the susceptibility. 

\begin{figure}[hb!]
    \centering
    \includegraphics[scale=0.8]{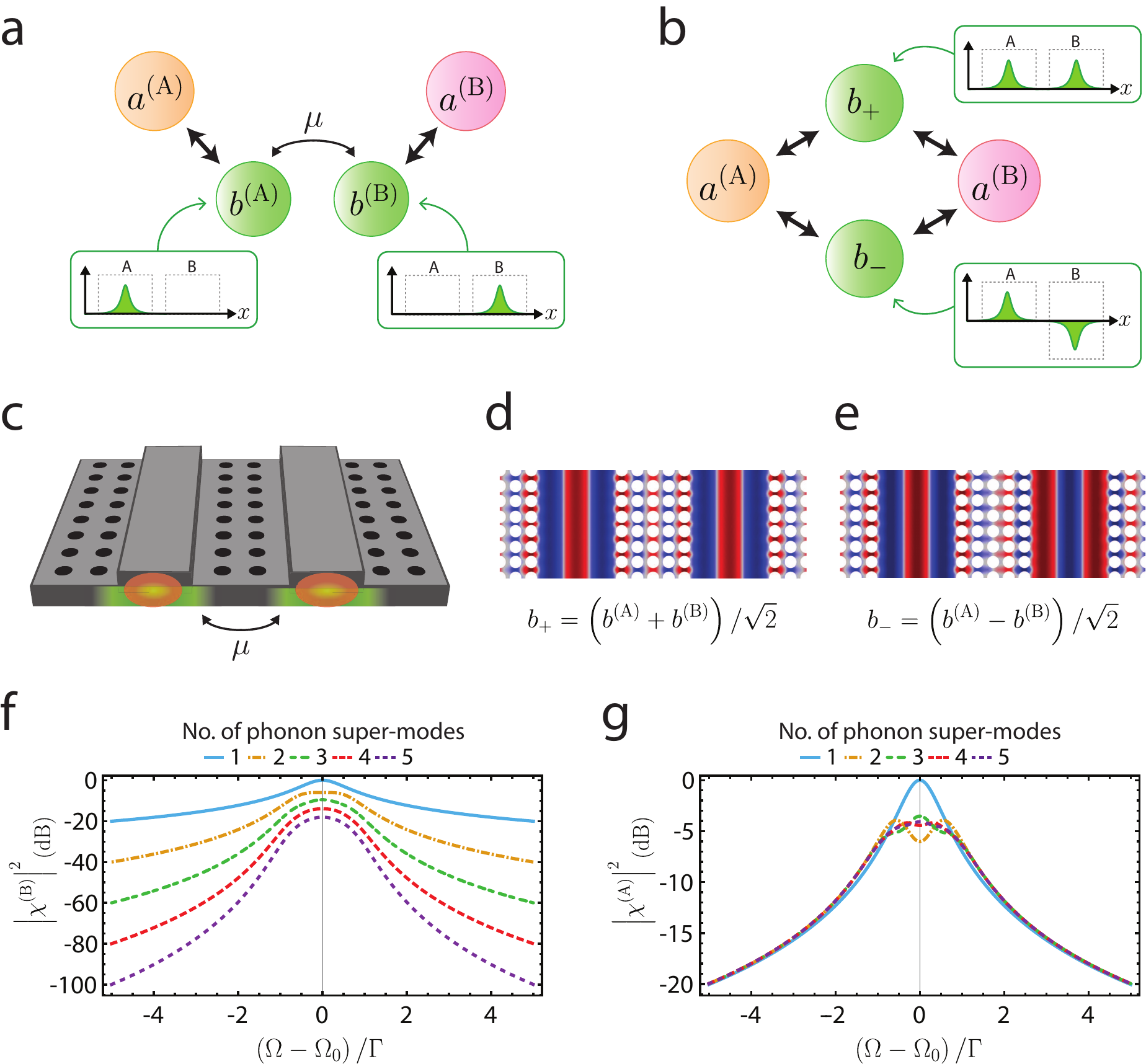}
    \caption{
    a. The optical field in each waveguide (A and B) is coupled to an acoustic mode through a FSBS process. The two spatial acoustic modes are coupled to each other with a rate \(\mu\).
    b. The same system can be described in the acoustic super-mode basis, using symmetric and anti-symmetric combinations of the spatial acoustic modes. The two acoustic super-modes are now decoupled, but the optical fields interact with both.
    c. Illustration of a device implementing acoustic coupling using a phononic crystal, such as was demonstrated in Ref. \cite{shin2015control}. The evanescent acoustic waves in the phononic crystal region enable coupling of the acoustic spatial modes.
    d. FEM simulation of the symmetric acoustic eigen-mode in a device such as illustrated in (c). 
    e. The anti-symmetric acoustic eigen-mode of the same geometry. 
    f. Calculated frequency response of the joint-susceptibility in waveguide B ($|\chi^\text{(B)}|^2$) of a coupled system of multiple acoustic modes (calculated using $\mu = \Gamma/2$).
    g. Calculated frequency response of the local susceptibility in waveguide A ($|\chi^\text{(A)}|^2$) of a coupled system of multiple acoustic modes (calculated using $\mu = \Gamma/2$).
    }
    \label{fig:multi_membrane}
\end{figure}

To explore this extended system, we consider two identical, spatially-separated waveguides, each guiding an optical mode and each supporting a phonon mode with a resonant frequency \(\Omega_0\) with an amplitude $b^\text{(A)}$, $b^\text{(B)}$ for waveguides A and B respectively. These two waveguides are designed such that they are optically decoupled but the two acoustic fields are coupled with a rate $\mu$, as illustrated in Fig. \ref{fig:multi_membrane}(a). An example of such a device, implementing the acoustic coupling using a phononic crystal design, was demonstrated in Ref. \cite{shin2015control} and is illustrated in Fig. \ref{fig:multi_membrane}(c). 
In this scheme, the acoustic modes are guided using line-defects in the the phononic crystal structure while the central region between the waveguides acts as a Bragg-reflector with finite reflectivity that enables evanescent coupling of the two phonon modes.
The acoustic coupling gives rise to symmetric and anti-symmetric acoustic super-modes, denoted \(b_+\) and \(b_-\), which are linear combinations of the spatial modes and have frequencies determined by the acoustic coupling rate \(\Omega_\pm = \Omega_0 \mp \mu\). 
This is demonstrated by a finite-element-method (FEM) simulation, solving for the acoustic eigen-modes of such a geometry, shown in Figs. \ref{fig:multi_membrane}(d) and \ref{fig:multi_membrane}(e). The two super-modes are now decoupled, as they are eigen-modes of the system, but are both coupled to the optical modes in both waveguides, as illustrated in Fig. \ref{fig:multi_membrane}(b).

Following the derivation presented in Appendix \ref{subsec:app_two-membrane}, the equations of motion are now given by
\begin{equation}
\begin{split}
b_{\pm} &= -i\left(\frac{1}{i(\Omega_\pm - \Omega) + \Gamma/2} \right) \sum_n\left({g_{\pm}^\text{(A)}}^* a_{n}^\text{(A)} {a_{n-1}^\text{(A)}}^{\dagger}+{g_{\pm}^\text{(B)}}^* a_{n}^\text{(B)} {a_{n-1}^\text{(B)}}^{\dagger} \right),\\
\frac{\partial {a}_{n}^\text{(A)}}{\partial z} &= -\frac{i}{v^\text{(A)}} \Big( g^\text{(A)}_{+} {a}_{n-1}^\text{(A)} b_{+} + {g_{+}^\text{(A)}}^* {a}_{n+1}^\text{(A)} b^{\dagger}_{+} + g^\text{(A)}_{-} {a}_{n-1}^\text{(A)} b_{-} + {g_{-}^\text{(A)}}^* {a}_{n+1}^\text{(A)} b^{\dagger}_{-} \Big),\\
\frac{\partial {a}_{n}^\text{(B)}}{\partial z} &= -\frac{i}{v^\text{(B)}} \Big( g_{+}^\text{(B)} {a}_{n-1}^\text{(B)} b_{+} + {g_{+}^\text{(B)}}^* {a}_{n+1}^\text{(B)} b^{\dagger}_{+} + g_{-}^\text{(B)} {a}_{n-1}^\text{(B)} b_{-} + {g_{-}^\text{(B)}}^* {a}_{n+1}^\text{(B)} b^{\dagger}_{-} \Big),
\end{split}
\label{eq:FSBS_two_membranes}
\end{equation}
where $b_{\pm} = \left(b^\text{(A)} \pm b^\text{(B)} \right)/\sqrt{2}$ are the amplitudes of the acoustic super-modes with frequencies $\Omega_{\pm}$, and $g_{\pm}^{(l)}$ is the Brillouin coupling rate of acoustic mode $b_{\pm}$ to the optical field in waveguide $l$ (for $l=$ \{A,B\}). By projecting the `\(\pm\)' super-modes on waveguides A and B we see that \( g^\text{(A)}_{\pm} = g/\sqrt{2}\), and \( g^\text{(B)}_{\pm} = \pm g/\sqrt{2}\), aresult of the symmetries of the super-modes. Since we assume the waveguides are identical, we also set the group velocities to the same value \(v = v^\text{(A)} = v^\text{(B)}\), yielding
\begin{equation}
    \begin{split}
    b_{\pm} &= -\frac{i}{\sqrt{2}} \left(\frac{1}{i(\Omega_\pm - \Omega) + \Gamma/2} \right) g^* \sum_n\left( a_{n}^\text{(A)} {a_{n-1}^\text{(A)}}^{\dagger} \pm a_{n}^\text{(B)} {a_{n-1}^\text{(B)}}^{\dagger} \right),\\
    \frac{\partial {a}_{n}^\text{(A)}}{\partial z} &= -\frac{i}{\sqrt{2}} \frac{1}{v} \Big( g {a}_{n-1}^\text{(A)} b_{+} + g^* {a}_{n+1}^\text{(A)} b^{\dagger}_{+} + g {a}_{n-1}^\text{(A)} b_{-} + g^* {a}_{n+1}^\text{(A)} b^{\dagger}_{-} \Big),\\
    \frac{\partial {a}_{n}^\text{(B)}}{\partial z} &= -\frac{i}{\sqrt{2}} \frac{1}{v} \Big( g {a}_{n-1}^\text{(B)} b_{+} + g^* {a}_{n+1}^\text{(B)} b^{\dagger}_{+} - g {a}_{n-1}^\text{(B)} b_{-} - g^* {a}_{n+1}^\text{(B)} b^{\dagger}_{-} \Big).
    \label{eq:eom_2_mem}
    \end{split}
\end{equation}

These equations have the same form as we saw earlier in Eq. (\ref{eq:phonon_field}), such that \(\partial_z b_{\pm} =0\), and the phonon super-modes are only dependent on the initial conditions of the optical fields. We assume again the optical inputs to be two tones in waveguide A, separated in frequency by \(\Omega\), and a single tone in waveguide B giving us
\begin{equation}
        b_{\pm} = -\frac{i}{\sqrt{2}}\left(\frac{1}{i\Delta_{\pm}+\Gamma/2} \right) g^* a_0^\text{(A)}(0) \  {a_{-1}^\text{(A)}}^{\dagger}(0),
\end{equation}
where we use the notation \( \Delta_{\pm} = \Omega_{\pm}-\Omega \).
Plugging the phonon fields into the optical equations of motion yields
\begin{equation}
    \begin{split}
    \frac{\partial {a}_{n}^\text{(A)}}{\partial z} &= - \frac{|g|^2}{v}  \left|{a_{-1}^\text{(A)}}^{\dagger}(0) \  a_0^\text{(A)}(0)\right| \left|\chi^\text{(A)}\right| \Bigg( {a}_{n-1}^\text{(A)} e^{i\phi^\text{(A)}}e^{i\Lambda} - {a}_{n+1}^\text{(A)} e^{-i\phi^\text{(A)}}e^{-i\Lambda} \Bigg),\\
    \frac{\partial {a}_{n}^\text{(B)}}{\partial z} &= - \frac{|g|^2}{v}  \left|{a_{-1}^\text{(A)}}^{\dagger}(0) \  a_0^\text{(A)}(0)\right| \left|\chi^\text{(B)}\right| \Bigg( {a}_{n-1}^\text{(B)} e^{i\phi^\text{(B)}}e^{i\Lambda} - {a}_{n+1}^\text{(B)} e^{-i\phi^\text{(B)}}e^{-i\Lambda} \Bigg),
    \end{split}
    \label{eq:EOM_with_chi}
\end{equation}
where we have defined the frequency response in each waveguide 
$
\chi^\text{(A)} = (1/2)[(i\Delta_{+}+\Gamma/2)^{-1} + (i\Delta_{-}+\Gamma/2)^{-1}] 
$
and 
$
\chi^\text{(B)} = (1/2)[(i\Delta_{+}+\Gamma/2)^{-1} - (i\Delta_{-}+\Gamma/2)^{-1}]$, with phases \(\phi^\text{(A)}=\arg(\chi^\text{(A)})\), \(\phi^\text{(B)}=\arg(\chi^\text{(B)})\), and the phase difference of the input tones \(\Lambda = \arg({a_{-1}^\text{(A)}}^{\dagger}(0) \  a_0^\text{(A)}(0))\).

The joint-susceptibility induced in waveguide B by the optical fields in waveguide A ($\chi^\text{(B)}$) is dramatically changed in the multi-phonon case, a result of the phase difference between the two complex terms. 
As demonstrated in Fig. \ref{fig:multi_membrane}(f), the coupling of multiple acoustic fields yields a high-order frequency response, showing a sharp frequency roll-off. 
This can also be seen by the interference of the two phonon super-modes, shown in Fig. \ref{fig:multi_membrane}(b), yielding a multi-pole function.
The susceptibility in waveguide A is also slightly altered around the acoustic resonance, as seen in Fig. \ref{fig:multi_membrane}(g), but decays as a Lorentzian away from the center frequency, similar to the single-phonon case.

Solving the equations, as described in the previous section yields the field amplitude at the output of waveguide B
\begin{equation}
    s^\text{(B)}_\text{out}(t) = \sqrt{P^\text{(B)}_0} \ e^{-i\omega^\text{(B)}_0 t} \ \exp \left[i \frac{\Gamma}{2}|\chi^\text{(B)}| \ G_B \sqrt{P_0^\text{(A)} P_{-1}^\text{(A)}} \ z \sin{\Big(\Omega t - \left( \phi + \Lambda\right)\Big)} \right],
    \label{eq:multi_mem_PM_output}
\end{equation}
which now has the modified frequency response $\chi^\text{(B)}$.

\vspace{5 mm}
This coupling of acoustic modes can be further extended to take into account an arbitrary number of phonons taking part in the transduction. For \(N\) identical waveguides, each supporting an acoustic mode with resonant frequency $\Omega_0$ and a nearest neighbour coupling with rate \(|\mu|e^{i\theta}\), the joint-sucseptibility frequency response is given by
\begin{equation}
    \chi^\text{(B)} = \sum_{m=1}^N V^{(N)}_{m} {V^{(1)}_{m}}^* \left(\frac{1}{i\Delta_m+\Gamma/2}\right),
    \label{eq:chi_B}
\end{equation}
where the detuning term for the \(m^\text{th}\) phonon super-mode is
%
$   
\Delta_m = ({\Omega}_0-\Omega) + 2|\mu| \cos[\pi m /(N+1) ],
$
%
and the phonon super-modes coefficients are given by
%
$
V_{m}^{(l)} = \sqrt{2/(N+1)}  \sin[\pi m l/(N+1) ]e^{i m \theta}
$,
%
described in more detail in Appendix \ref{subsec:app_Multi-membrane}.
The response described by Eq. (\ref{eq:chi_B}) yields sharper frequency responses with the addition of coupled acoustic modes, as demonstrated in Fig. \ref{fig:multi_membrane}(f).

\subsection*{\label{subsec:noise}Spontaneaous scattering}
The Brillouin coupling in the system results in scattering of the tone in waveguide B even in the absence of input fields into waveguide A.  
This spontaneous scattering is a result of the thermal occupation of the acoustic modes taking part in the Brillouin process \cite{boyd_book,kharel2016_Hamiltonian}.
This can also be understood as the noise associated with the loss of the acoustic mode, through the fluctuation-dissipation theorem \cite{raymer1981stimulated,boyd1990noise}. 
For practical consideration, when utilizing nonlocal susceptibility to transduce information \cite{shin2015control,kittlaus2018rf,kittlaus2018non,diamandi2017opto}, it is essential to understand how this noise is imparted from the elastic field onto the optical fields.

Assuming a single acoustic mode with frequency $\Omega_0$, an optical tone at frequency $\omega_0$ with amplitude $a_{0}$ will be scattered to sidebands at frequencies \(\omega_0 \pm \Omega_0\). The amplitudes of the spontaneously scattered light are given by \cite{boyd1990noise,kharel2016_Hamiltonian}
\begin{equation}
    \begin{split}
        a_{-1}(z,\tau) &= -i\frac{g^*}{v}a_{0} \int_0^\tau d\tau'\int_0^z dz'\eta^\dagger(z',\tau') e^{-\frac{\Gamma}{2} (\tau-\tau')},\\
        a_{1}(z,\tau) &= -i\frac{g}{v}a_{0} \int_0^\tau d\tau'\int_0^z dz'\eta(z',\tau') e^{-\frac{\Gamma}{2} (\tau-\tau')},
    \end{split}
    \label{eq:noise_EOM}
\end{equation}
where \(\eta(z,t)\) is the Langevin force corresponding to the phonon dissipation rate, with statistics \(\langle \eta(z,t)\rangle = 0\) and \(\langle \eta^\dagger(z,t)\eta(z',t')\rangle = \bar{n}\Gamma \delta(z-z')\delta(t-t')\) when evaluating the ensemble averages \cite{kharel2016_Hamiltonian}. \(\bar{n}\) is the average number of thermally occupied phonons following a Bose-Einstein distribution \(\bar{n} = \left[\exp \left( \hbar \Omega / k_\text{B} T \right)-1 \right]^{-1}\), with \(T\) denoting the temperature, and \(k_\text{B}\) the Boltzmann constant. These equations apply for both waveguides A and B.

The spectral density of the spontaneous scattering is given by \cite{kharel2016_Hamiltonian}
\begin{equation}
    S(\omega) = \hbar \omega_0 G_\text{B} P z \left(\frac{\Gamma}{2}\right)^2 \left( \frac{\bar{n}+1}{\left(\omega - \left(\omega_0 - \Omega_0\right)\right)^2+(\Gamma/2)^2} + \frac{\bar{n}}{\left(\omega - \left(\omega_0 + \Omega_0\right)\right)^2+(\Gamma/2)^2}\right),
    \label{eq:noise_spect}
\end{equation}
showing a Lorentzian lineshape with a full-width-at half-maximum \(\Gamma\).
In the case of coupled acoustic modes, as discussed in the previous section, Eq. (\ref{eq:noise_EOM}) can be generalized
\begin{equation}
    \begin{split}
    a^{(l)}_{-1}(z,\tau) &= -i\frac{g^*}{v}a_{0} \sum_{m=1}^{N}{ {V^{(l)}_{m}}^* \int_{0}^{\tau}{ d\tau'\int_{0}^{z} {dz'\eta_m^\dagger(z',\tau') e^{-\frac{\Gamma}{2} (\tau-\tau')}}}},\\
    a^{(l)}_{1}(z,\tau) &= -i\frac{g}{v}a_{0} \sum_{m=1}^{N}{ {V^{(l)}_{m}} \int_{0}^{\tau}{ d\tau'\int_{0}^{z} {dz'\eta_m(z',\tau') e^{-\frac{\Gamma}{2}(\tau-\tau')}}}},
    \end{split}
    \label{eq:noise_EOM_N_membranes}
\end{equation}
where \(\eta_m\) is the Langevin force associated with the \(m^\text{th}\) phonon super-mode, and \(V^{(l)}_{m}\) is the same coefficient from Eq. (\ref{eq:chi_B}). Since the acoustic super-modes are orthogonal, we use the fact that the thermal phonons in each eigen-mode are uncorrelated and follow \(\langle \eta_m(z,t)\rangle = 0\) and \(\langle \eta_m^\dagger(z,t)\eta_{m'}(z',t')\rangle = \bar{n}_{m}\Gamma \delta_{m,m'}\delta(z-z')\delta(t-t')\). The spectral density is now given by
\begin{equation}
    S(\omega) = \hbar \omega_0 G_\text{B} P z \left(\frac{\Gamma}{2}\right)^2 \sum_{m=1}^N |V^{(l)}_{m}|^2 \left( \frac{\bar{n}_m+1}{\left(\omega - \left(\omega_0 - \Omega_m\right)\right)^2+(\Gamma/2)^2} + \frac{\bar{n}_{m}}{\left(\omega - \left(\omega_0 + \Omega_m\right)\right)^2+(\Gamma/2)^2}\right),
    \label{eq:noise_spect_N_membranes}
\end{equation}
where \(\bar{n}_{m}\) is the average thermal phonon occupation of the \(m^\text{th}\) phonon super mode. We see that the spontaneous scattering is a sum of Lorentzian lineshapes, and does not show a multi-pole like the one observed for a transduced signal. In the limit of very weak coupling between the phonon modes, the frequency differences of the super-modes will be small and we we can approximate \(\bar{n}_{m} \approx \bar{n} \) and \(\omega_0 \pm \Omega_m \approx \omega_0 \pm \Omega_0\). Using \(\sum_{m=1}^N |V^{(l)}_{m}|^2 =1\) we see that the spontaneous spectrum is consistent with the one obtained from a single phonon device, given by Eq. (\ref{eq:noise_spect}), exhibiting a Lorentzian frequency response.

Interestingly, the contribution of the thermal scattering does not affect the driven coherent acoustic field, as it only adds phase fluctuations, without changing the coherent generation of phonons. This can be shown from Eq. (\ref{eq:a}) where the contribution to the phonon field is the sum of terms 
$
({a^\text{(A)}_{-1}}^\dagger a^\text{(A)}_{0} + {a^\text{(A)}_{0}}^\dagger a^\text{(A)}_{1} + {a^\text{(B)}_{-1}}^\dagger a^\text{(B)}_{0} + {a^\text{(B)}_{0}}^\dagger a^\text{(B)}_{1})
$
which equals zero when plugging in the terms from Eq. (\ref{eq:noise_EOM}).

\section{\label{sec:discussion}Discussion and conclusion}
The theoretical model presented in this work describes the nonlinear dynamics of a class of opto-mechanical forward-Brillouin active devices, utilizing the nonlocal nature of the acoustic modes taking part in the process. We have shown that the optical beat-note of the light in waveguide A induces a nonlinear response in the spatially separated waveguide B. In the absence of optical dispersion, this results in a phase modulation, where the modulation depth is set by the Brillouin gain, acoustic properties, interaction length, and optical powers. This joint-susceptibility is enabled by the long lifetime of the phonons taking part in the interaction, allowing them to propagate a large distance spanning multiple optical wavelengths, and connecting optically decoupled regions of the device.

We have shown that this nonlocal susceptibility can be mediated by multiple coupled phonons. In this case, the phonons can be treated as acoustic super-modes, i.e. the eigen-modes of the coupled system, which are all interacting with the optical fields. The coherent interference of these acoustic super-modes yields a joint-susceptibility exhibiting a multi-pole frequency response, giving rise to a sharp frequency roll-off and high out-of-band suppression. This drastically altered frequency response is unique to the joint-susceptibility where the phonon super-modes interfere destructively. In contrast, for a process where all fields of interest are contained in the same waveguide (i.e. in waveguide A), the acousto-optic forcing function drives all acoustic modes in phase, yielding a frequency response that decays as a single-pole Lorentzian. When analyzing spontaneous Brillouin scattering, the response will also exhibit an overall Lorentzain lineshape with slightly different spectral features near the resonant frequencies.

In our analysis, we have considered a constant optical group velocity for all optical tones, such that the optical group-velocity dispersion (GVD) is zero. This is a good approximation in many systems \cite{kang2009tightly,kharel2016_Hamiltonian,kittlaus2016_FSBS} and can further be avoided through the design of zero-GVD waveguides \cite{petrov2004zero,zhao2012wideband}. When considering non-zero GVD, the modulation experienced by the light in waveguide B will no longer be purely phase modulated and will exhibit some residual intensity modulation. A numerical investigation of the effect of GVD is presented in Appendix \ref{subsec:app_dispersion}, showing that for GVD values of common materials and waveguide designs the residual amplitude modulation is five orders of magnitude smaller than the transduced signal. 

Brillouin nonlocal interactions can be extended to include multiple spatial optical modes through use of an inter-band scattering processes \cite{kang2010all,kittlaus2017_Intermodal}. In this scenario, the acoustic modes that mediate the acousto-optic interaction have a non-vanishing axial wave-vector yet are still mostly transverse in nature. The dynamics of this nonlocal inter-modal interaction result in power transfer between two optical spatial modes in each waveguide, unlike the phase modulation we saw for the single mode case. This scheme has been recently experimentally implemented, demonstrating a wide-band nonreciprocal modulator \cite{kittlaus2018non}. Further details of the inter-modal joint-susceptibility are discussed in Appendix \ref{subsec:app_XMPPER} and in Ref. \cite{kittlaus2018non}.

Tailorable, nonlocal, nonlinear responses are useful in many practical applications, such as filtering, coherent signal addition, opto-acoustic storage and spectral analysis \cite{eggleton2019brillouin}. Further, the immunity of the forcing function which generates the acoustic field in the input waveguide to additive thermal noise in the device, enables these devices to be cascaded in series without suffering reduced performance. This could have further applications such as spectral analysis, channelizing and sensing. While we specifically focus on phonon mediated non-local interactions, these same coupled wave dynamics can apply to any three wave processes with similar properties.

\section{\label{sec:acknowledgments}Acknowledgments}
This research was supported primarily by the Packard Fellowship for Science and Engineering.
We also acknowledge funding support from ONR YIP (N00014-17-1-2514).
N.T.O acknowledges support from the National Science Foundation Graduate Research Fellowship under Grant DGE1122492.
Part of the research was carried out at the Jet Propulsion Laboratory, California Institute of Technology, under a contract with the National Aeronautics and Space Administration.

\appendix

\section{\label{subsec:app_Hamiltonian}Deriving the equations of motion}
We start our analysis by considerng a system of two spatially separated optical waveguides, denoted A and B, both coupled to a single acoustic mode. We can separate the Hamiltonian of the system to terms describing the optical fields in waveguides A and B, the acoustic field, and the acousto-optic interaction in each waveguide
\begin{equation}
H = H_{\text{opt}}^\text{(A)} + H_{\text{opt}}^\text{(B)} + H_{\text{ac}} + H_{\text{int}}^\text{(A)} + H_{\text{int}}^\text{(B)}.
\label{eq:H_parts}
\end{equation}
Following the treatment of Refs. \cite{sipe2016hamiltonian,kharel2016_Hamiltonian} the different terms are given by
\begin{equation}
\begin{split}
H_{\text{opt}}^{\text{(A)}} &= \sum_{n}\hbar\int{dz\ {a_{n}^\text{(A)}}^\dagger(z)\hat{\omega}_{n}^\text{(A)} a_{n}^\text{(A)}(z)}, \\
H_{\text{opt}}^{\text{(B)}} &= \sum_{n}\hbar\int{dz\ {a_{n}^\text{(B)}}^{\dagger}(z)\hat{\omega}_{n}^\text{(B)} a_{n}^\text{(B)}(z)}, \\
H_{\text{ac}} &= \hbar\int{dz\ b^{\dagger}(z)\hat{\Omega}_0 b(z)}, \\
H_{\text{int}}^{\text{(A)}} &= \sum_{n}\hbar\int{dz\ {g_{n}^\text{(A)}}^* a_{n}^\text{(A)}(z) {a_{n-1}^\text{(A)}}^\dagger(z) b^{\dagger}(z) e^{-i\Delta k_{n}^{\text{(A)}}z}} + \text{H.C.},\\
H_{\text{int}}^{\text{(B)}} &= \sum_{n}\hbar\int{dz\ {g_{n}^\text{(B)}}^* a_{n}^\text{(B)}(z) {a_{n-1}^\text{(B)}}^\dagger(z) b^{\dagger}(z) e^{-i\Delta k_{n}^{\text{(B)}}z}} + \text{H.C.}, 
\end{split}
\label{eq:H_parts_def}
\end{equation}
where \(\Delta k_{n}^{\text{(A)}}=q_0-(k_{n}^{\text{(A)}}-k_{n-1}^{\text{(A)}})\) and \(\Delta k_{n}^{\text{(B)}}=q_0-(k_{n}^{\text{(B)}}-k_{n-1}^{\text{(B)}})\) are the phase mismatch between the phonon and the photons in each of the two waveguides. We assume each waveguide supports a single optical spatial mode, and the index \(n\) sums over all optical tones coupled through the acousto-optic interaction. The coupling rate \(g\) can have both photo-elastic and radiation pressure components, which can be evaluated following Ref. \cite{kharel2016_Hamiltonian}.

We can calculate the dynamics of the acoustic field using the Heisenberg equation
\begin{equation}
    \begin{split}
    \dot{a}_n^\text{(A)}(t)&=\frac{1}{i\hbar}\left[{a}_n^\text{(A)},H\right],
    \end{split}
    \qquad
    \begin{split}
    \dot{a}_n^\text{(B)}(t)&=\frac{1}{i\hbar}\left[{a}_n^\text{(B)},H\right],
    \end{split}
    \qquad
    \begin{split}
    \dot{b}(t)&=\frac{1}{i\hbar}\left[b,H\right],
    \end{split}
    \label{eq:Heisenberg_eq}
\end{equation}
and using the commutation relations
\begin{equation}
    \begin{split}
    &\left[{a}_n^\text{(A)}(z,t),{a_m^\text{(A)}}^{\dagger}(z',t)\right]=\delta\left(z-z'\right)\delta_{n,m},\\ &\left[{a}_n^\text{(B)}(z,t),{a_m^\text{(B)}}^{\dagger}(z',t)\right]=\delta\left(z-z'\right)\delta_{n,m},
    \end{split}
    \qquad
    \begin{split}
    &\left[{a}_n^\text{(A)}(z,t),{a_m^\text{(B)}}^{\dagger}(z',t)\right]=0,\\ &\left[b(z,t),b^{\dagger}(z',t)\right]=\delta\left(z-z'\right),
    \end{split}
    \label{eq:commutator}
\end{equation}
yields
\begin{equation}
\begin{split}
\dot{a}_{n}^\text{(A)}(z,t)&=-i\hat{\omega}_n a_{n}^\text{(A)}(z,t)-i\left(g_{n}^\text{(A)} a^\text{(A)}_{n-1}(z,t)b(z,t)e^{i\Delta k_{n}^{\text{(A)}}z}+{{g^\text{(A)}}^*_{n+1}} a^\text{(A)}_{n+1}(z,t) b^{\dagger}(z,t) e^{-i\Delta k_{n+1}^{\text{(A)}}z}\right),\\
\dot{a}_{n}^\text{(B)}(z,t)&=-i\hat{\omega}_n a_{n}^\text{(B)}(z,t)-i\left(g_{n}^\text{(B)} a_{n-1}^\text{(B)}(z,t)b(z,t)e^{i\Delta k_{n}^{\text{(B)}}z}+g_{n+1}^\text{(B)*} a_{n+1}^\text{(B)}(z,t) b^{\dagger}(z,t) e^{-i\Delta k_{n+1}^{\text{(B)}}z}\right),\\
\dot{b}(z,t)&=-i\hat{\Omega}_0b(z,t)-i\sum_{n}\left({g_{n}^\text{(A)}}^* a_{n}^\text{(A)}(z,t) {a^\text{(A)}_{n-1}}^{\dagger}(z,t) e^{-i\Delta k_{n}^{\text{(A)}}z}+g_{n}^\text{(B)*} a_{n}^\text{(B)}(z,t) {a_{n-1}^\text{(B)}}^{\dagger}(z,t) e^{-i\Delta k_{n}^{\text{(B)}}z}\right).
\end{split}
\label{eq:EOM_w_disp_opertor}
\end{equation}

We keep the dispersion operators to first order for the optical and acoustic fields, \( \hat{\omega}_n \approx \omega_{n}-i{v}_n\partial_z\), \(
\hat{\Omega}_0 \approx\Omega_0-i{v}_\text{ac}\partial_z\), and add phonon dissipation by including an imaginary part to the acoustic frequency, \( \Omega_0 \rightarrow\Omega_0-i\left(\Gamma/2\right) \). Rewriting the equation in the rotating frame by factoring out the fast oscillating term
\begin{equation}
\begin{split}
a_n(z,t)&\rightarrow a_n(z,t) e^{-i\omega_n t},
\end{split}
\qquad
\begin{split}
b(z,t)&\rightarrow b(z,t) e^{-i\Omega t},
\end{split}
\end{equation}
where \(\Omega=\omega_n-\omega_{n-1}\), we now have
\begin{equation}
\begin{split}
\dot{a}_{n}^\text{(A)}+{v}_n^\text{(A)}\frac{\partial {a}_{n}^\text{(A)}}{\partial z}&=-i\left(g_{n}^\text{(A)} a^\text{(A)}_{n-1} b e^{i\Delta {k_{n}^\text{(A)}}z}+{g_{n+1}^\text{(A)}}^* a^\text{(A)}_{n+1} b^{\dagger} e^{-i\Delta {k_{n+1}^\text{(A)}}z}\right),\\
\dot{a}_{n}^\text{(B)}+{v}_n^\text{(B)}\frac{\partial {a}_{n}^\text{(B)}}{\partial z}&=-i\left(g_{n}^\text{(B)} a_{n-1}^\text{(B)} b e^{i\Delta {k_{n}^\text{(B)}}z}+{g_{n+1}^\text{(B)}}^* a_{n+1}^\text{(B)} b^{\dagger} e^{-i\Delta {k^\text{(B)}_{n+1}}z}\right),\\
\dot{b}+{v}_\text{ac}\frac{\partial b}{\partial z} + i\left(\Omega_0-\Omega-i \Gamma/2\right)b&=-i\sum_{n}\left({g_{n}^\text{(A)}}^* a_{n}^\text{(A)} {a^\text{(A)}_{n-1}}^{\dagger} e^{-i\Delta {k^\text{(A)}_n}z} +{g_{n}^\text{(B)}}^* a_{n}^\text{(B)} {a^\text{(B)}_{n-1}}^{\dagger} e^{-i\Delta {k^\text{(B)}_n}z}\right).
\end{split}
\label{eq:FSBS}
\end{equation}

In the case of forward Brillouin scattering (FSBS) the phonon field \(b(z)\) is close to its cut-off frequency, with a small axial wave-vector and vanishing group velocity \cite{kharel2016_Hamiltonian}, such that we can set  \({v}_\text{ac}\rightarrow0\). We further assume that the optical mode has a constant group velocity in the frequency range of interest for each of the two waveguides, equivalent to no optical group velocity dispersion (GVD), such that
\({v}_n = {v}\) and \(\Delta k_{n} = \Delta k = 0\) \cite{rakich2010tailoring,kharel2016_Hamiltonian}. We also set the opto-mechanical coupling rates to be equal for all optical frequencies, \(g_n = g\). The steady state phonon field envelope now has the form
\begin{equation}
b=-i\left(\frac{1}{i\Delta + \Gamma/2}\right)\sum_n\left({g^\text{(A)}}^* a_{n}^\text{(A)} {a^\text{(A)}_{n-1}}^{\dagger}+{g^\text{(B)}}^* a_{n}^\text{(B)} {a_{n-1}^\text{(B)}}^{\dagger} \right),
\label{eq:b}
\end{equation}
where \(\Delta = \Omega_0-\Omega\), and the optical fields envelopes are given by
\begin{equation}
    \begin{split}
        \frac{\partial {a}_{n}^\text{(A)}}{\partial z}&=-\frac{i}{{v}^\text{(A)}}\left(g^\text{(A)} a^\text{(A)}_{n-1} b +{g^\text{(A)}}^* a^\text{(A)}_{n+1} b^{\dagger} \right),\\
        \frac{\partial {a}_{n}^\text{(B)}}{\partial z}&=-\frac{i}{{v}^\text{(B)}}\left(g^\text{(B)} a_{n-1}^\text{(B)} b +{g^\text{(B)}}^* a_{n+1}^\text{(B)} b^{\dagger} \right).
    \end{split}
    \label{eq:a_app}
\end{equation}
Calculating the spatial derivative of the phonon field in Eq. (\ref{eq:b}) results in
\begin{equation}
    \frac{\partial b}{\partial z} = -i\left(\frac{1}{i\Delta + \Gamma/2}\right) \sum_n \left( {g^\text{(A)}}^* \frac{\partial a_{n}^\text{(A)}}{\partial z}  {a^\text{(A)}_{n-1}}^{\dagger} + {g^\text{(A)}}^* a_{n}^\text{(A)} \frac{\partial {a^\text{(A)}_{n-1}}^{\dagger} }{\partial z} + {g^\text{(B)}}^* \frac{\partial a_{n}^\text{(B)}}{\partial z}  {a_{n-1}^\text{(B)}}^{\dagger} + {g^\text{(B)}}^* a_{n}^\text{(B)} \frac{\partial {a_{n-1}^\text{(B)}}^{\dagger}}{\partial z}\right),\\
    \label{eq:b_deriv}
\end{equation}
and plugging in Eq. (\ref{eq:a_app}) yields
\begin{equation}
\begin{split}
    \frac{\partial b}{\partial z} = -\left(\frac{1}{i\Delta + \Gamma/2}\right) \sum_n \Bigg[ 
    &{g^\text{(A)}}^*\Bigg(g^\text{(A)} b \left( \left|a^\text{(A)}_{n-1}\right|^2 - \left|a_{n}^\text{(A)}\right|^2\right) + {g^\text{(A)}}^* b^{\dagger} \left( a^\text{(A)}_{n+1} {a^\text{(A)}_{n-1}}^{\dagger} - a_{n}^\text{(A)} {a^\text{(A)}_{n-2}}^{\dagger}\right) \Bigg) + \\
    &{g^\text{(B)}}^*\Bigg(g^\text{(B)} b \left( \left|a_{n-1}^\text{(B)}\right|^2 - \left|a_{n}^\text{(B)}\right|^2\right) + {g^\text{(B)}}^* b^{\dagger} \left( a_{n+1}^\text{(B)} {a_{n-1}^\text{(B)}}^{\dagger} - a_{n}^\text{(B)} {a^\text{(B)}_{n-2}}^{\dagger}\right) \Bigg)\Bigg] = 0 .
    \end{split}
    \label{eq:b_deriv_plug}
\end{equation}
We can see directly from the above equation that the spatial derivative vanishes when we sum over all \(n\) for which the field operators are non-zero. Hence the phonon field is constant in space and can be determined by its value at \(z=0\)
\begin{equation}
    b = -i\left(\frac{1}{i\Delta + \Gamma/2}\right)\sum_{n} \left({g^\text{(A)}}^* {a^\text{(A)}_{n}}^{\dagger} {a}^\text{(A)}_{n+1} + {g^\text{(B)}}^* {a^\text{(B)}_n}^{\dagger} {a}^\text{(B)}_{n+1} \right) \biggr\rvert_{z=0} .
    \label{eq:phonon_field_app}
\end{equation}

\begin{figure}
    \centering
    \includegraphics[scale=0.8]{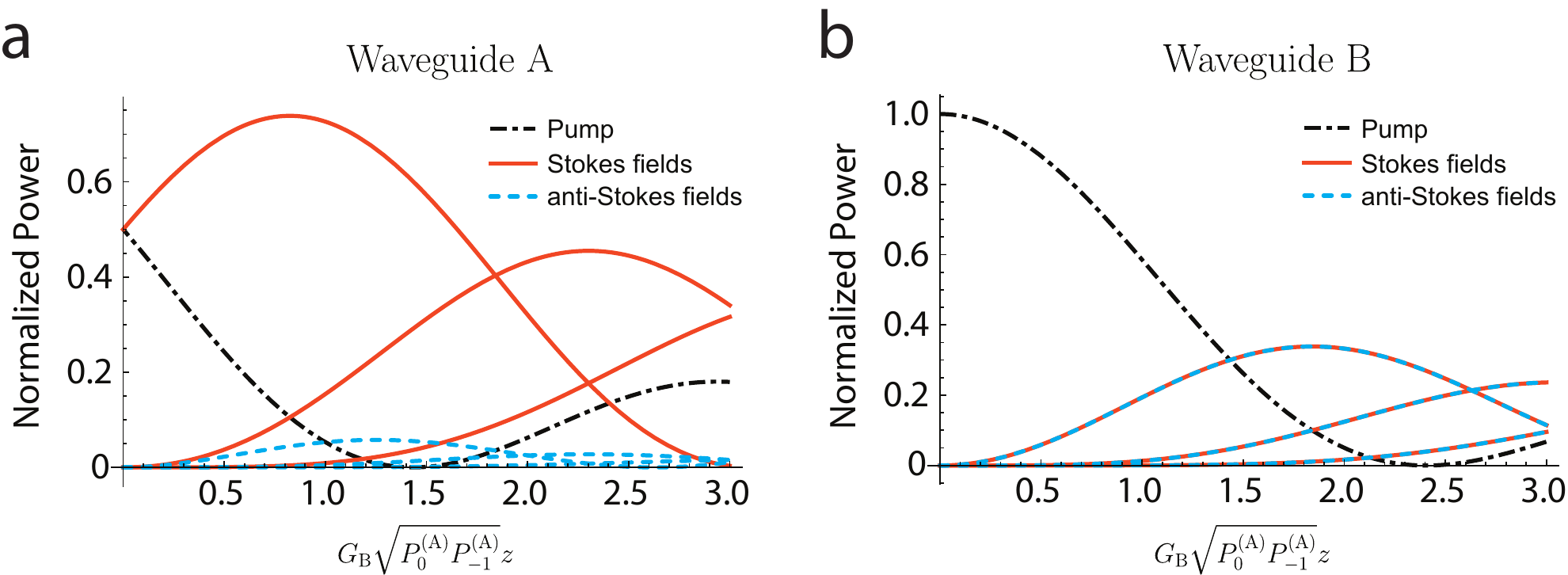}
    \caption{
    a. Normalized power in the different optical tones propagating in waveguide A, given a two-tone input. The pump tone ($a_0$) is shown in black, Stokes ($a_n, n<0$) tones in red, and anti-Stokes ($a_n, n>0$) tones in blue.
    b. Normalized power in the different optical tones propagating in waveguide B, given a single tone input modulated by the acoustic field.
    }
    \label{fig:App_Bessel}
\end{figure}
%


\section{\label{subsec:app_emit}Dynamics in the `emit' waveguide}
Restating Eq. (\ref{eq:a4}), we saw that the equations of motion describing the tones in waveguide A, which has a pump and a Stokes wave at its input, are given by
\begin{equation}
    a_n^\text{}(z) =  a^\text{}_{0}(0) J_{-n} \left(\xi z \right)e^{i\left(\phi+\Lambda \right)n} + a^\text{}_{-1}(0) J_{-(n+1)} \left(\xi z \right) e^{i\left(\phi+\Lambda \right)(n+1)},
    \label{eq:FSBS_emit}
\end{equation}
where \( \xi  = \sqrt{P_0^\text{} P_{-1}^\text{}} G_\text{B} (\Gamma/2) |\chi|\), \( \chi = \left({i(\Omega_0 - \Omega)+\Gamma/2}\right)^{-1} \), \(\phi = \arg (\chi ) \), and \(\Lambda = \arg ({a^\text{}_{-1}}^{\dagger}(0) \ {a}^\text{}_{0}(0) ) \). The total field amplitude propagating through the waveguide is the sum of all the amplitudes, $s(z,t) \propto \sum_n {a}_n(z) e^{-i (\omega_0 + n\Omega )t}$, yielding
\begin{equation}
    s(z,t) \propto e^{-i \omega_0 t} \left({a}_{0}(0) \sum_n J_{-n}(\xi z) e^{i(\phi+\Lambda)n} e^{-i n \Omega t} + {a}_{-1}(0) \sum_n J_{-(n+1)}(\xi z) e^{i(\phi+\Lambda)(n+1)} e^{-i n \Omega t} \right).
\end{equation}
By using the index substitutions $m=-n$ and $m'=-(n+1)$ in the two sums respectively, and rearranging the equation, we have
\begin{equation}
    s(z,t) \propto \Big( {a}_{0}(0) e^{-i \omega_0 t} + {a}_{-1}(0) e^{-i (\omega_0-\Omega) t} \Big)  \sum_{m'} J_{m'}(\xi z) e^{-i(\phi+\Lambda-\Omega t)m'},
\end{equation}
which can also be expressed as a phase modulation, by using the Jacobi-Anger expansion
\begin{equation}
    s(z,t) \propto \Big( {a}_{0}(0) e^{-i \omega_0 t} + {a}_{-1}(0) e^{-i (\omega_0-\Omega) t} \Big) \exp \left[i  \frac{\Gamma}{2} \left|\chi\right| G_\text{B} \sqrt{P_0^\text{} P_{-1}^\text{}} \ z \sin{\Big(\Omega t - \left( \phi + \Lambda\right)\Big)} \right],
\end{equation}
where we have plugged back the argument of the Bessel function. 

We see that the two input tones are phase modulated as they traverse the device, with the modulation index increasing linearly with the propagation length. In contrast, the intensity profile is unaffected by the forward-Brillouin process. Examining the power as would be measured by a photo-detector yields
\begin{equation}
     \left|s(0,t)\right|^2 \propto \left|a_{0}(0)\right|^2 + \left|a_{-1}(0)\right|^2 + a_{0}(0) {a_{-1}}^\dagger(0) e^{-i \Omega t} + {a_{0}}^\dagger(0) a_{-1}(0) e^{i \Omega t},
     \label{eq:beat_0}
\end{equation}
and in terms of optical powers
\begin{equation}
     \left|s(0,t)\right|^2 = P_0 + P_{-1} + 2 \sqrt{P_0 P_{-1}} \cos{\left(\Omega t - \Lambda\right)},
\end{equation}
revealing the DC terms and the oscillation at the beat-note frequency $\Omega$, the frequency separation between the two tones. We can see directly that the propagation through the device and the Brillouin scattering that takes place does not affect the intensity envelope.

\subsection*{\label{subsec:FSBS}Small-signal forward Brillouin gain}
Using the equations we have derived, we can also examine the gain experienced by a small Stokes signal as it propagates through the device along with a pump wave. Using Eq. (\ref{eq:a4}), which assumes pump and Stokes waves at its input, we examine the Stokes ($n=-1$) field amplitude
\begin{equation}
    a_{-1}(z) =  a_{0}(0) J_{1} \left(\xi z \right)e^{-i\left(\phi+\Lambda \right)} + a_{-1}(0) J_{0} \left(\xi z \right),
    \label{eq:FSBS_Stokes}
\end{equation}
where \( \xi  = \sqrt{P_0(0) P_{-1}(0)} G_\text{B} (\Gamma/2) |\chi|\), \(\phi = \arg (\chi ) \), and \(\Lambda = \arg ({a^\text{}_{-1}}^{\dagger}(0) \ {a}^\text{}_{0}(0) ) \).
Rearranging the equation, we have
\begin{equation}
    \begin{split}
        a_{-1}^\text{}(z) &= a^\text{}_{-1}(0) \left( \left|\frac{a^\text{}_{0}(0)}{a^\text{}_{-1}(0)}\right| J_{1} \left(\xi z \right)e^{-i\phi} + J_{0} \left(\xi z \right)\right),
    \end{split}
    \label{eq:FSBS_2}
\end{equation}
and in terms of the optical powers, $P_n \propto |a_n|^2 $, this yields
\begin{equation}
    \begin{split}
        P_{-1}^\text{}(z) &= P^\text{}_{-1}(0) \left| \sqrt{\frac{P^\text{}_{0}(0)}{P^\text{}_{-1}(0)}} J_{1} \left(\xi z \right)e^{-i\phi} + J_{0} \left(\xi z \right)\right|^2.
    \end{split}
    \label{eq:FSBS_3}
\end{equation}

We now take the small signal limit, where we assume the pump power is much larger than the input signal and pump depletion can be neglected, equivalent to $\left( \xi z \right) \ll 1$. We can expand the Bessel functions to first order, $J_0(x)\approx1$, $J_1(x)\approx x/2$, yielding
\begin{equation}
    \begin{split}
        P_{-1}^\text{}(z) &= P^\text{}_{-1}(0) \left| 1+\frac{1}{2} G_\text{B} P_0^\text{} \frac{\Gamma}{2} \chi\left(\Omega\right) z \right|^2,
    \end{split}
    \label{eq:FSBS_4}
\end{equation}
showing the quadratic growth of power in the Stokes tone in the small signal limit, consistent with the results derived in Ref. \cite{kharel2016_Hamiltonian}.


\section{\label{subsec:app_two-membrane}Two coupled phonons}

We now consider a device where two identical separate Brillouin active devices, denoted A and B, each supporting an optical mode \(a(z)\) and an acoustic mode \(b(z)\), where the acoustic modes have a resonant frequency $\Omega_0$ and are coupled with a rate \(\mu\). The acoustic Hamiltonian of Eq. (\ref{eq:H_parts_def}) now has the form
\begin{equation}
    H_{\text{ac}} = \hbar \int{ dz\ \left( {b^\text{(A)}}^{\dagger}(z)\hat{\Omega}_0 b^\text{(A)}(z) + {b^\text{(B)}}^{\dagger}(z)\hat{\Omega}_0 b^\text{(B)}(z) - {b^\text{(A)}}^{\dagger}(z) \ \mu \ b^\text{(B)}(z) - b^\text{(A)}(z)\ \mu \ {b^\text{(B)}}^{\dagger}(z) \right)},
    \label{eq:H_ac_coupled}
\end{equation}
which can also be written in matrix form
\begin{equation}
    H_{\text{ac}} = \hbar\int{dz\ \begin{pmatrix} {b^\text{(A)}}^{\dagger} & {b^\text{(B)}}^{\dagger} \end{pmatrix} \begin{pmatrix} \hat{\Omega}_0 & -\mu \\ -\mu & \hat{\Omega}_0 \end{pmatrix} \begin{pmatrix} b^\text{(A)} \\ b^\text{(B)} \end{pmatrix}}.
    \label{eq:H_ac_matrix_form}
\end{equation}

Assuming phase matching conditions are met, the interaction Hamiltonian terms are now 
\begin{equation}
    \begin{split}
        H_{\text{int}}^{\text{(A)}} &= \sum_{n}\hbar\int{dz\ {g_{n}^\text{(A)}}^* a_{n}^\text{(A)}(z) {a_{n-1}^\text{(A)}}^{\dagger}(z) {b^\text{(A)}}^{\dagger}(z) } + \text{H.C.},\\
        H_{\text{int}}^{\text{(B)}} &= \sum_{n}\hbar\int{dz\ {g_{n}^\text{(B)}}^* a_{n}^\text{(B)}(z) {a_{n-1}^\text{(B)}}^{\dagger}(z) {b^\text{(B)}}^{\dagger}(z)} + \text{H.C.}, 
    \end{split}
\end{equation}
where we see the three wave process in each waveguide, between two optical tones and an acoustic mode. The optical Hamiltonian terms will stay unchanged from Eq. (\ref{eq:H_parts_def}). We can now calculate the equations of motion of the phonons using Eq. (\ref{eq:Heisenberg_eq}), yielding
\begin{equation}
    \begin{split}
    \dot{b}^\text{(A)}(z,t)&=-i\hat{\Omega}_0 b^\text{(A)}(z,t) + i \mu b^\text{(B)}(z,t) - i\sum_{n} {g_{n}^\text{(A)}}^*  a_{n}^\text{(A)}(z,t) {a_{n-1}^\text{(A)}}^{\dagger}(z,t),\\
    \dot{b}^\text{(B)}(z,t)&=-i\hat{\Omega}_0 b^\text{(B)}(z,t) + i \mu b^\text{(A)}(z,t) - i\sum_{n} {g_{n}^\text{(B)}}^*  a_{n}^\text{(B)}(z,t) {a_{n-1}^\text{(B)}}^{\dagger}(z,t).
    \end{split}
\end{equation}

This result is consistent with a temporal coupled-mode theory approach such as described in Refs. \cite{little1997microring,shin2015control}. Alternatively, we can diagonalize Eq. (\ref{eq:H_ac_matrix_form}) to the eigen-basis such that we have two decoupled phonon modes
\begin{equation}
H_{\text{ac}} = \hbar\int{dz\ \begin{pmatrix} b^{\dagger}_{+} & b^{\dagger}_{-} \end{pmatrix} \begin{pmatrix} \hat{\Omega}_+ & 0 \\ 0 & \hat{\Omega}_- \end{pmatrix} \begin{pmatrix} b_{+} \\ b_{-} \end{pmatrix}}.
\end{equation}
These phonon `super-modes' extend spatially to both waveguides, and can be written as a superposition of the phonon modes
\(b_{\pm} = \left(b^\text{(A)} \pm b^\text{(B)}\right)/\sqrt{2},\) and their respective frequencies \(\hat{\Omega}_{\pm} = \hat{\Omega}_{0}\mp \mu\), which retain the commutation relations \( [b_{\pm}(z,t),b_{\pm}^{\dagger}(z',t)]=\delta\left(z-z'\right)\) and \([b_{\pm}(z,t),b_{\mp}^{\dagger}(z',t)] = 0\). The interaction Hamiltonian terms in this basis are now
\begin{equation}
    \begin{split}
        H_{\text{int}}^{\text{(A)}} &= \sum_{n}\hbar\int{dz\ \left( {g_{+}^\text{(A)}}^* a_{n}^\text{(A)}(z) {a_{n-1}^\text{(A)}}^{\dagger}(z) b_+^{\dagger}(z) + {g_{-}^\text{(A)}}^* a_{n}^\text{(A)}(z) {a_{n-1}^\text{(A)}}^{\dagger}(z) b_-^{\dagger}(z)\right)} + \text{H.C.},\\
        H_{\text{int}}^{\text{(B)}} &= \sum_{n}\hbar\int{dz\ \left( {g_{+}^\text{(B)}}^* a_{n}^\text{(B)}(z) {a_{n-1}^\text{(B)}}^{\dagger}(z) b_+^{\dagger}(z) + {g_{-}^\text{(B)}}^* a_{n}^\text{(B)}(z) {a_{n-1}^\text{(B)}}^{\dagger}(z) b_-^{\dagger}(z)\right)} + \text{H.C.},
    \end{split}
\end{equation}
where \({g^{(l)}_{\pm}}\) is the coupling rate between the phonon super-modes and the optical tones in waveguide $l$, where $l=$ \{A,B\}.

We can calculate the equations of motion using Eq. (\ref{eq:Heisenberg_eq}), and keep the dispersion operator to first order for the optical modes, and to zero order for the acoustic modes which have a vanishing group velocity. Additionally, we factor out the fast oscillating term, and add a dissipation rate \(\Gamma\) to the phonon modes
\begin{equation}
\begin{split}
\dot{a}_{n}^\text{(A)} + {v}\frac{\partial {a}_{n}^\text{(A)}}{\partial z} &= -i \Big( g^\text{(A)}_{+} {a}_{n-1}^\text{(A)} b_{+} + {g_{+}^\text{(A)}}^* {a}_{n+1}^\text{(A)} b^{\dagger}_{+} + g^\text{(A)}_{-} {a}_{n-1}^\text{(A)} b_{-} + {g_{-}^\text{(A)}}^* {a}_{n+1}^\text{(A)} b^{\dagger}_{-} \Big),\\
\dot{a}_{n}^\text{(B)} + {v}\frac{\partial {a}_{n}^\text{(B)}}{\partial z} &= -i \Big( g^\text{(B)}_{+} {a}_{n-1}^\text{(B)} b_{+} + {g_{+}^\text{(B)}}^* {a}_{n+1}^\text{(B)} b^{\dagger}_{+} + g^\text{(B)}_{-} {a}_{n-1}^\text{(B)} b_{-} + {g_{-}^\text{(B)}}^* {a}_{n+1}^\text{(B)} b^{\dagger}_{-} \Big),\\
\dot{b}_{\pm} + i \left( \Omega_{\pm}-\Omega - i \frac{\Gamma}{2} \right) b_{\pm} &=  -i \sum_n\left({g^\text{(A)}}^*_{\pm} a_{n}^\text{(A)} {a_{n-1}^\text{(A)}}^{\dagger}+{g^\text{(B)}}^*_{\pm} a_{n}^\text{(B)} {a_{n-1}^\text{(B)}}^{\dagger} \right).
\end{split}
\label{eq:FSBS_two_membranes_app}
\end{equation}

\section{\label{subsec:app_Multi-membrane}Generalizing to multiple coupled phonons}
Using the same approach, we can expand the previous derivation to an arbitrary number of acoustic coupled modes. Assuming $N$ identical waveguides supporting the same phonon resonance and nearest neighbour coupling, the acoustic Hamiltonian is now given by
\begin{equation}
    H_{\text{ac}} = \hbar\int{dz\ \begin{pmatrix} {b^\text{(A)}}^{\dagger} & {b^\text{(B)}}^{\dagger} & \cdots & {b^{(N)}}^{\dagger} \end{pmatrix} 
    \begin{pmatrix} \hat{\Omega}_0 & \mu & \\
    \mu^* & \hat{\Omega}_0 & \mu \\
    & \ddots & \ddots & \ddots \\
    & &\mu^* & \hat{\Omega}_0 & \mu \\ 
    & & & \mu^* & \hat{\Omega}_0 \end{pmatrix} 
    \begin{pmatrix} b^\text{(A)} \\ b^\text{(B)} \\ \vdots \\ b^{(N)} \end{pmatrix}},
    \label{eq:N_membrane_Hamiltonian}
\end{equation}
where \(\mu = |\mu|e^{i\theta}\), and we assume \(0\) in all matrix elements not on the three main diagonals. The interaction Hamiltonian in the \(l^\text{th}\) waveguide is now given by
\begin{equation}
    H_{\text{int}}^{(l)} = \sum_{n}\hbar\int{dz\  {g}^* a_{n}^{(l)}(z) {a_{n-1}^{(l)}}^{\dagger}(z) {b^{(l)}}^{\dagger}(z) } + \text{H.C.}
\end{equation}

The tri-diagonal matrix  of Eq. (\ref{eq:N_membrane_Hamiltonian}) can be diagonalized, yielding \(N\) distinct eigenvalues \cite{noschese2013tridiagonal}
\begin{equation}
    \hat{\Omega}_m = \hat{\Omega}_0 + 2\left|\mu\right| \cos\left({\frac{\pi m}{N+1}}\right),
    \label{eq:Omega_m_app}
\end{equation}
and the phonon fields can be decomposed into phonon eigen-modes
\begin{equation}
    \begin{split}
    b^{(l)} = \sum_{m=1}^N V^{(l)}_{m} \ b_m,
    \end{split}
    \qquad
    \begin{split}
    V^{(l)}_{m} = \sqrt{\frac{2}{N+1}} \sin\left({\frac{\pi m l}{N+1}}\right)e^{i m \theta},
    \end{split}
    \label{eq:Vnm}
\end{equation}
where we have chosen the notation such that \(m\) is an index for phonon super-modes, \(l\) enumerates the different waveguides, and \(n\) sums the different optical tones spaced by frequency \(\Omega\). We can now rewrite the acoustic and interaction terms of the Hamiltonian
\begin{equation}
    \begin{split}
    H_{\text{ac}} &= \sum_{m=1}^N \hbar\int{dz\ {b^{\dagger}_m} \hat{\Omega}_m {b_m}},\\
    H_{\text{int}}^{(l)} &= \sum_{m=1}^N \sum_{n}\hbar\int{dz\  {g_{m}^{(l)}}^* a_{n}^{(l)}(z) {a_{n-1}^{(l)}}^{\dagger}(z) b_m^{\dagger}(z) } + \text{H.C.}
    \end{split}
\end{equation}
where the rate \({g^{(l)}_{m}}\), denoting the coupling of the \(m^\text{th}\) phonon eigen-modes to the optical tones in the \(l^\text{th}\) waveguide, can be described in terms of the single-phonon coupling  \(g^{(l)}_{m} = g {V^{(l)}_{m}}\) using Eq. (\ref{eq:Vnm}).

Calculating the equations of motion using Eq. (\ref{eq:Heisenberg_eq}), under the same assumptions as in the previous section, we can write the field envelopes for the phonon eigen-modes and the optical fields in the \(l^\text{th}\) membrane
\begin{equation}
    \begin{split}
    b_{m} &= \ -i\left(\frac{1}{i\Delta_m+\Gamma_m/2} \right) \sum_n \sum_{l=1}^N {g^{(l)}_{m}}^* a^{(l)}_{n} {a^{(l)}_{n-1}}^{\dagger},\\
    \frac{\partial{a}^{(l)}_{n}}{\partial z} &= -\frac{i}{v} \sum_{m=1}^N \Big( g^{(l)}_{m} a^{(l)}_{n-1} b_{m} + {g^{(l)}_{m}}^* a^{(l)}_{n+1} b^{\dagger}_{m}
    \Big).
    \label{eq:eom_N_mem}
    \end{split}
\end{equation}

Assuming a two tone input into waveguide A with a frequency spacing \(\Omega\), and a single tone in the input of waveguide B, the phonon field from Eq. (\ref{eq:eom_N_mem}) is now 
\begin{equation}
    b_{m} = -i\left(\frac{1}{i\Delta_m+\Gamma_m/2} \right) {g_{m}^\text{(A)}}^* a^\text{(A)}_{0}(0) \  {a^\text{(A)}_{-1}}^{\dagger}(0),
\end{equation}
where \(\Delta_m = \Omega_m - \Omega\) and \(\Gamma_m\) are the detuning and the loss of the \(m^{\text{th}}\) phonon eigen-mode respectively.
Following the same steps as described in Section \ref{subsec:multi_membrane}, we can find the equation of motion for the optical tones in waveguide \(l\)
\begin{equation}
    \frac{\partial{a}^{(l)}_{n}}{\partial z} = - \frac{1}{v} \left|g\right|^2 \left|{a_{-1}^\text{(A)}}^{\dagger}(0) \  a_0^\text{(A)}(0)\right| \left|\chi^{(l)}\right| \Bigg( a^{(l)}_{n-1} e^{i\phi^{(l)}}e^{i\Lambda} - a^{(l)}_{n+1} e^{-i\phi^{(l)}}e^{-i\Lambda} \Bigg),
\end{equation}
where the frequency response is denoted \(\chi^{(l)} = \sum_{m=1}^N V^{(l)}_{m} {V^\text{(A)}_{m}}^* (i\Delta_m+\Gamma_m/2)^{-1}\), \(\phi^{(l)} = \arg(\chi^{(l)})\), and the phase difference between the two input tones to waveguide A is given by \(\Lambda = \arg({a_{-1}^\text{(A)}}^{\dagger}(0) \  a_0^\text{(A)}(0))\). 
This equation has an identical form to Eq. (\ref{eq:reccurance_with_phases}), and following the steps in Section \ref{sec:theory} we can find the optical field envelopes
\begin{equation}
    \begin{split}
        a_n^\text{(A)}(z) &=  a^\text{(A)}_{0}(0) J_{-n} \left(\xi^\text{(A)} z \right)e^{i\left(\phi^\text{(A)}+\Lambda \right)n} + a^\text{(A)}_{-1}(0) J_{-(n+1)} \left(\xi^\text{(A)} z \right) e^{i\left(\phi^\text{(A)}+\Lambda \right)(n+1)},\\
        a_n^\text{(B)}(z) &=  a_{0}^\text{(B)}(0) J_{-n} \left(\xi^\text{(B)} z \right)e^{i\left(\phi^\text{(B)}+\Lambda \right)n},
    \end{split}
    \label{eq:a4_app}
\end{equation}
where \(\xi^\text{(A)} = (\Gamma/2)|\chi^\text{(A)}|G_\text{B} (P_0^\text{(A)} P_{-1}^\text{(A)})^{1/2} z\) and \(\xi^\text{(B)} = (\Gamma/2)|\chi^\text{(B)}|G_\text{B} (P_0^\text{(A)} P_{-1}^\text{(A)})^{1/2} z\). 

We note that waveguides A and B do not have to be in specific spatial positions along the coupled-waveguide array, and are defined only by the optical inputs. The field amplitude at the output of waveguide B is now
\begin{equation}
    s^\text{(B)}_\text{out}(t) = \sqrt{P_{0}^\text{(B)}} e^{-i \omega^\text{(B)}_0 t}\sum_n J_{n} \left( \frac{\Gamma}{2}\left|\chi^\text{(B)}\right| \ G_\text{B} \sqrt{P_0^\text{(A)} P_{-1}^\text{(A)}} \ z \right) e^{-i \Big(\Omega t - \left( \phi^\text{(B)} + \Lambda\right) + \pi\Big)n},
\end{equation}
and by using the Jacobi-Anger expansion, this can also be expressed as
\begin{equation}
    s^\text{(B)}_\text{out}(t) = \sqrt{P_{0}^\text{(B)}} \ e^{-i\omega^\text{(B)}_0t} \ \exp \left[i\ \frac{\Gamma}{2}\left|\chi^\text{(B)}\right| \ G_\text{B} \sqrt{P_0^\text{(A)} P_{-1}^\text{(A)}} \ z \sin{\Big(\Omega t - \left( \phi^\text{(B)} + \Lambda\right)\Big)} \right].
\end{equation}

\begin{figure}
    \centering
    \includegraphics[scale=0.8]{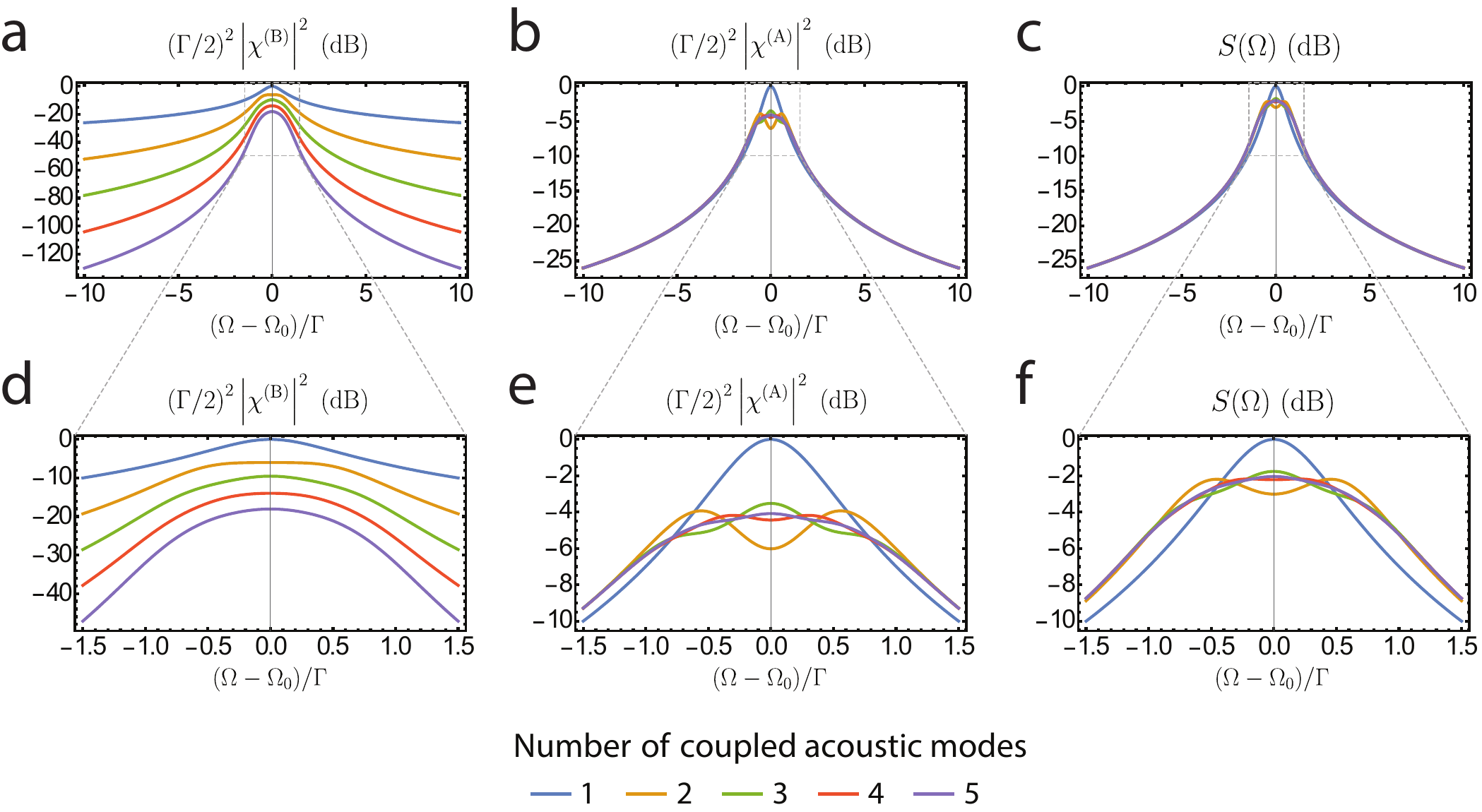}
    \caption{
    a. Frequency response of the `joint-susceptibility' in waveguide B for different numbers of coupled acoustic modes, following Eq. (\ref{eq:chi_B}), showing a sharper response with the addition of coupled phonon modes to the system.
    b. Brillouin induced susceptibility in waveguide A for different numbers of coupled acoustic modes. Now, the frequency response is modified around the resonance, but decays in a similar manner to a Lorentzain function.
    c. Normalized spectrum of a spontaneous-scattering sideband, for a different numbers of coupled acoustic modes, following Eq. (\ref{eq:noise_spect_N_membranes}).
    d-f. Magnified view of plots a-c, respectively.
    All of the simulated frequency responses were calculated using $\mu = \Gamma/2$.
    }
    \label{fig:App_freq}
\end{figure}

We can examine the frequency response at the output of waveguide B for a few cases:
\begin{itemize}
    \item 
    Single acoustic mode:
    \begin{equation}
        \chi^\text{(B)}= \frac{1}{\Omega-\left( \Omega_0-i\Gamma/2\right)}
    \end{equation}
    This is a single pole function, with a maximum amplitude response at \(\Omega = \Omega_0\).
    \item 
    Two coupled acoustic modes:
    \begin{equation}
        \chi^\text{(B)}= \frac{i\mu}{\left[ \Omega-\left( \Omega_+-i\Gamma/2\right)\right]\left[ \Omega-\left( \Omega_--i\Gamma/2\right)\right]}
    \end{equation}
    This function has two poles, with a maximum amplitude response at \(\Omega = \Omega_0 \pm \sqrt{\mu^2 - (\Gamma/2)^2}\).
    \item
    Three coupled acoustic modes:
    \begin{equation}
        \chi^\text{(B)}= \frac{-\mu^2 }{\left[ \Omega-\left( \Omega_+-i\Gamma/2\right)\right]\left[ \Omega-\left( \Omega_--i\Gamma/2\right)\right] \left[ \Omega-\left( \Omega_0-i\Gamma/2\right)\right]}
    \end{equation}
    Now we have a function with three poles, which will have maximum amplitude at frequecies \(\Omega = \Omega_0\), \(\Omega = \Omega_0 \pm \sqrt{\frac{4}{3}\mu^2 + \frac{2}{3}\mu\sqrt{\mu^2-6(\Gamma/2)^2}- (\Gamma/2)^2}\).
\end{itemize}
Fig. \ref{fig:App_freq}(d) illustrates some of these different frequency responses for different numbers of coupled acoustic modes. Calculating the frequency response at the output of waveguide A reveals that the functions \(\chi^\text{(A)}\) always have \(N-1\) zeros, such that the response is that of a single pole, decaying as a Lorentzian function, as is demonstrated in Fig. \ref{fig:App_freq}(c).

\section{\label{subsec:app_XMPPER}Limited optical cascading}
\begin{figure}
    \centering
    \includegraphics[scale=0.8]{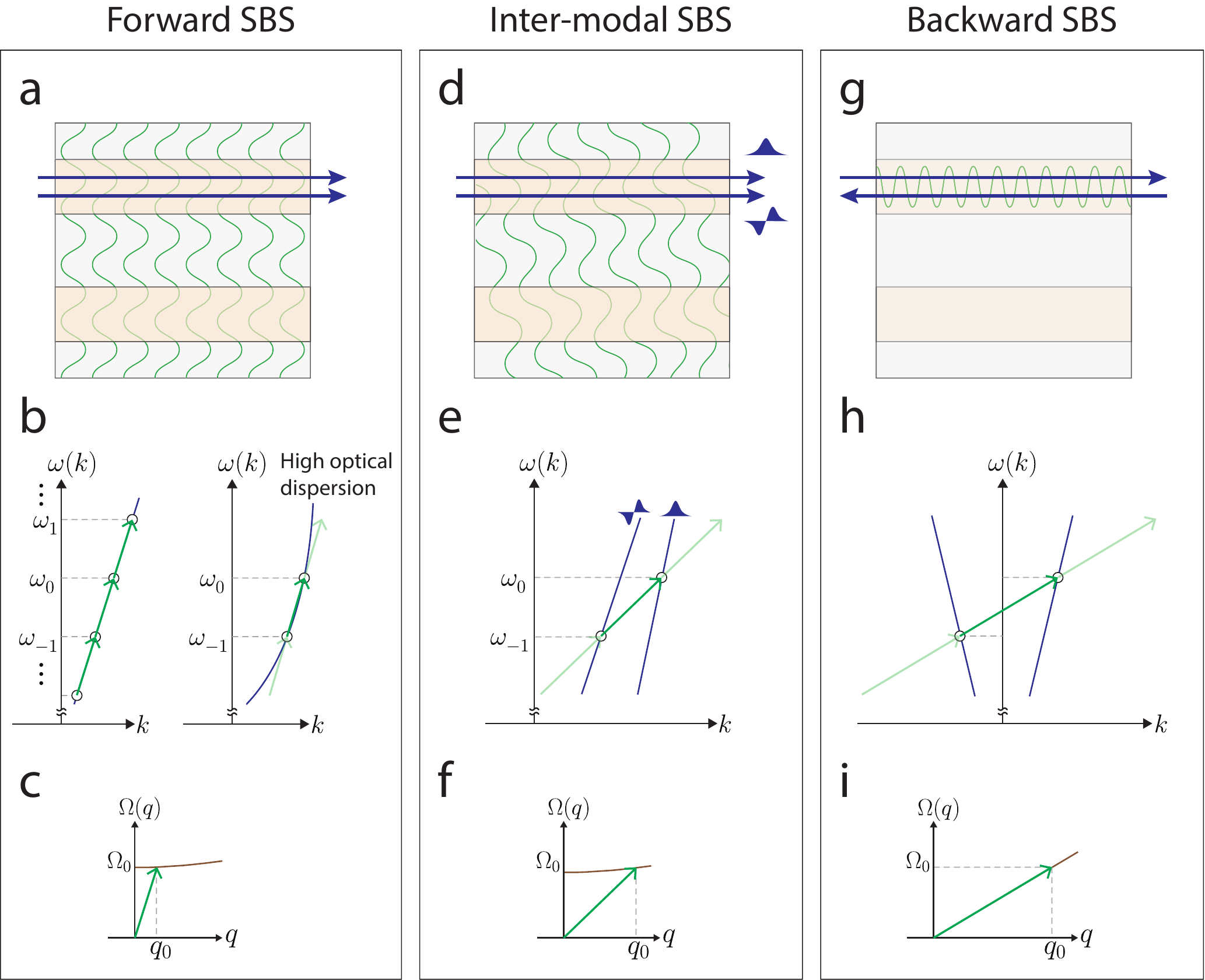}
    \caption{
    a. Forward SBS (stimulated Brillouin scattering), where the acoustic mode is perpendicular to the optical propagation, and can explore a large space and overlap with another waveguide.
    b. The dispersion curve of forward SBS shows the single optical mode taking part in the process. In the case of no optical dispersion (constant group velocity), a cascaded array of optical tones is possible. In the limit of high dispersion, the cascading is limited to a finite number of optical tones, determined by the phase matching conditions.
    c. The acoustic dispersion curve, illustrating the cutoff mode, with a vanishing axial wave-vector taking part in forward SBS.
    d. Inter-modal SBS, where two different spatial optical modes take part in the nonlinear process, and an acoustic mode which can extend far beyond the optical mode region.
    e. The optical dispersion curves show that the phonon is phase-matched to a transition between two different optical spatial modes.
    f. The acoustic mode taking part in the nonlinear process is a cutoff mode with a small axial wave-vector. This results in an acoustic mode with a non-zero axial wave-vector. 
    g. In backward SBS, the optical tones are counter-propagating. This geometry can be phased-matched to a bulk acoustic mode confined spatially to the same region as the optical waves, propagating along the same axis.
    h. The dispersion diagram of the backward SBS process is phase matched with the scattering between forward and backward propagating optical tones.
    i. In backward SBS, a bulk acoustic mode with a large axial wave-vector, propagating parallel to the optical waves, is phased-matched in this nonlinear process.
    }
    \label{fig:XMPPER}
\end{figure}

We now examine the case where there is no optical cascading, such that only two optical tones propagate in each of the waveguides A and B. This can be a result of high optical dispersion, leading to phase matching of the nonlinear process only between two tones \cite{kharel2016_Hamiltonian}, or when an inter-modal Brillouin process drives the acoustic field \cite{kang2010all,kittlaus2017_Intermodal}. The different geometries and conditions for Brillouin scattering, and the resulting phase matched nonlinear processes are illustrated in Fig. \ref{fig:XMPPER}.

Starting with Eq. (\ref{eq:a}), we consider two tones in waveguides A and B by keeping only $n=\{-1,0\}$, leaving us with
\begin{equation}
    \begin{split}
    \frac{\partial {a}^\text{(A)}_0}{\partial z} &= -\frac{i}{v^\text{(A)}} {g^\text{(A)}} b {a}^\text{(A)}_{-1} ,\\
    \frac{\partial {a}^\text{(A)}_{-1}}{\partial z} &= -\frac{i}{v^\text{(A)}} {g^\text{(A)}}^* b^{\dagger}{a}^\text{(A)}_{0} ,\\
    \end{split}
    \qquad
    \begin{split}
    \frac{\partial {a}^\text{(B)}_0}{\partial z} &= -\frac{i}{v^\text{(B)}} {g^\text{(B)}} b {a}^\text{(B)}_{-1} ,\\
    \frac{\partial {a}^\text{(B)}_{-1}}{\partial z} &= -\frac{i}{v^\text{(B)}} {g^\text{(B)}}^* b^{\dagger}{a}^\text{(B)}_{0},\\
    \end{split}
\end{equation}
and the acoustic field
\begin{equation}
    b = -i\left(\frac{1}{i\Delta + \Gamma/2}\right) \left({g^\text{(A)}}^* {a^{\dagger}}^\text{(A)}_{-1} {a}^\text{(A)}_{0} + {g^\text{(B)}}^* {a^{\dagger}}^\text{(B)}_{-1} {a}^\text{(B)}_{0} \right).
    \label{eq:XM_b}
\end{equation}

We note that in this inter-modal process, the phonons do not have a vanishing group velocity as in the FSBS case. However, the axial spatial evolution of the acoustic field is very slow compared to the optical fields, and can be adiabatically eliminated, such that Eq. (\ref{eq:XM_b}) is still valid \cite{otterstrom2018silicon,kittlaus2018non}.
In this case, the number of photons in each waveguide is conserved, which can be seen from the derivatives

\begin{equation}
    \frac{\partial}{\partial z} \left( {a^{\dagger}}^\text{(A)}_{0} {a}^\text{(A)}_0 + {a^{\dagger}}^\text{(A)}_{-1} {a}^\text{(A)}_{-1}\right) = 0,
    \qquad
     \frac{\partial}{\partial z} \left( {a^{\dagger}}^\text{(B)}_{0} {a}^\text{(B)}_0 + {a^{\dagger}}^\text{(B)}_{-1} {a}^\text{(B)}_{-1}\right) = 0.
\end{equation}

The phonon field is not constant in space, and is in fact amplified or damped along the propagation direction \cite{wolff2017cascaded}. However, assuming again two tones in the input to waveguide A and a single tone to waveguide B, to leading order (assuming $|{a^{\dagger}}^\text{(A)}_{-1} {a}^\text{(A)}_{0}|^2 >> |{a^{\dagger}}^\text{(B)}_{-1} {a}^\text{(B)}_{0}|^2$), we can drop the second term in Eq. (\ref{eq:XM_b}), yielding
\begin{equation}
    b = -i \chi {g^\text{(A)}}^* {a^{\dagger}}^\text{(A)}_{-1} {a}^\text{(A)}_{0},
\end{equation}
where we have again used the notation \( \chi = \left({i\Delta+\Gamma/2}\right)^{-1} \). We can use our notation for the nonlinear susceptibilities
\begin{equation}
    \begin{split}
    &\gamma^\text{(A)} = -\frac{i}{{v}^\text{(A)}} \left|g^\text{(A)}\right|^2 \chi^*,
    \end{split}
    \qquad
    \begin{split}
    &\gamma^\text{(B)} = -\frac{i}{{v}^\text{(B)}} g^\text{(B)*} g^\text{(A)} \chi^*,
    \end{split}
\end{equation}
giving us the equations of motion for the optical waves
\begin{equation}
    \begin{split}
    \frac{\partial {a}^\text{(A)}_0}{\partial z} &= i {\gamma^\text{(A)}}^* {a^{\dagger}}^\text{(A)}_{-1} {a}^\text{(A)}_{0} {a}^\text{(A)}_{-1} ,\\
    \frac{\partial {a}^\text{(A)}_{-1}}{\partial z} &= i \gamma^\text{(A)}  {a}^\text{(A)}_{-1} {a^{\dagger}}^\text{(A)}_{0} {a}^\text{(A)}_{0} ,\\
    \end{split}
    \qquad
    \begin{split}
    \frac{\partial {a}^\text{(B)}_0}{\partial z} &=i {\gamma^\text{(B)}}^*  {a^{\dagger}}^\text{(A)}_{-1} {a}^\text{(A)}_{0} {a}^\text{(B)}_{-1} ,\\
    \frac{\partial {a}^\text{(B)}_{-1}}{\partial z} &=i {\gamma^\text{(B)}} {a}^\text{(A)}_{-1} {a^{\dagger}}^\text{(A)}_{0} {a}^\text{(B)}_{0},
    \end{split}
\end{equation}
illustrating the nonlocal nonlinear susceptibility in the two-tone case. A complete description of the spatial dynamics of this case can be found in Ref. \cite{kittlaus2018non}.

For completeness, we mention that a backwards-Brillouin process, where the two optical modes are counter propagating, require a phonon with a large wave-vector to take part in the process \cite{boyd_book}. This is commonly achieved by using a bulk-acoustic mode, as illustrated in Fig. \ref{fig:XMPPER}(i). In this case, the acoustic field has a similar spatial extent as the optical waves. However, if a cutoff guided acoustic mode takes part in the three wave process, it could extend in space further than the overlap region, similar to the inter-modal SBS case.

\section{\label{subsec:app_dispersion}Effects of dispersion}
In the presence of optical group velocity dispersion (GVD), different optical frequencies will have different group velocities. When using different optical wavelengths in waveguides A and B, this can lead to a phase mismatch between the phonons driven in waveguide A and those needed for efficient phase modulation in waveguide B. This phase mismatch can be expressed in terms of the driving frequency \(\Omega\) and the frequency dependent group velocity \(v\left(\omega\right)\) 
\begin{equation}
    \Delta q = \Omega \left(\frac{1}{v\left(\omega^\text{(A)}_0\right)} - \frac{1}{v\left(\omega^\text{(B)}_0\right)}\right).
\end{equation}
where we assume the optical dispersion is identical in both waveguides A and B. For the phase mismatch to be small over the device length, i.e. $\Delta q L < \pi/2$, this gives the condition
\begin{equation}
    \left(\frac{1}{v\left(\omega^\text{(A)}_0\right)} - \frac{1}{v\left(\omega^\text{(B)}_0\right)}\right) < \frac{\pi}{2 \Omega L}.
\end{equation}

Another consideration to take into account in the presence of GVD is that the cascaded optical tones will not be phased matched for all scattering orders. This will result in modulation of the light in waveguide B which is not purely a phase modulation \cite{wolff2017cascaded}. We numerically demonstrate the effect of GVD by keeping the second order term of the dispersion operator in Eq. (\ref{eq:EOM_w_disp_opertor}) for the acoustic field, and the optical fields in both waveguides A and B
\begin{equation}
\begin{split}
& b =-i\left(\frac{1}{i\Delta + \Gamma/2}\right)  \sum_{n}\left(g^*_{n} a^\text{(A)}_{n} {a^\text{(A)}_{n-1}}^{\dagger} e^{-i\Delta k_{n}z}+g^*_{n} a^\text{(B)}_{n} {a^\text{(B)}_{n-1}}^{\dagger} e^{-i\Delta k_{n}z}\right),\\
& -i \left(\frac{\partial^2 \omega}{\partial k^2}\right)\frac{\partial^2 {a}_{n}}{\partial z^2} +{v}_n\frac{\partial {a}_{n}}{\partial z} =-i\left(g_{n} a_{n-1} b e^{i\Delta k_{n}z}+g^*_{n+1} a_{n+1} b^{\dagger} e^{-i\Delta k_{n+1}z}\right).
\end{split}
\label{eq:EOM_w_GVD}
\end{equation}
where we now have frequency dependent optical group velocity \(v_{n} = \partial_k \omega\bigr\rvert_{\omega=\left(\omega_0+n\Omega\right)}\) and a phase mismatch term \(\Delta k_n = q_0-\left(k_n-k_{n-1}\right) \approx \Omega\left(v_{0}^{-1} - v_{n}^{-1}  \right)\), where we have used a piece-wise linear approximation. The dispersion leads to non-symmetric sideband around the carrier and now the optical field is not only phase modulated, but also has residual amplitude modulation (RAM). We quantify the RAM by looking at the components of the intensity at frequency \(\Omega\), normalized to the total scattered power
\begin{equation}
    \text{RAM} = \frac{\sum_n \left( {a_{n-1}^\text{(B)}}^{\dagger} a_n^\text{(B)} + a_{n-1}^\text{(B)} {a_n^\text{(B)}}^{\dagger}\right)}{\sum_{n \neq 0} a_n^\text{(B)} {a_n^\text{(B)}}^{\dagger}}.
\end{equation}

\begin{figure}
    \centering
    \includegraphics[scale=0.8]{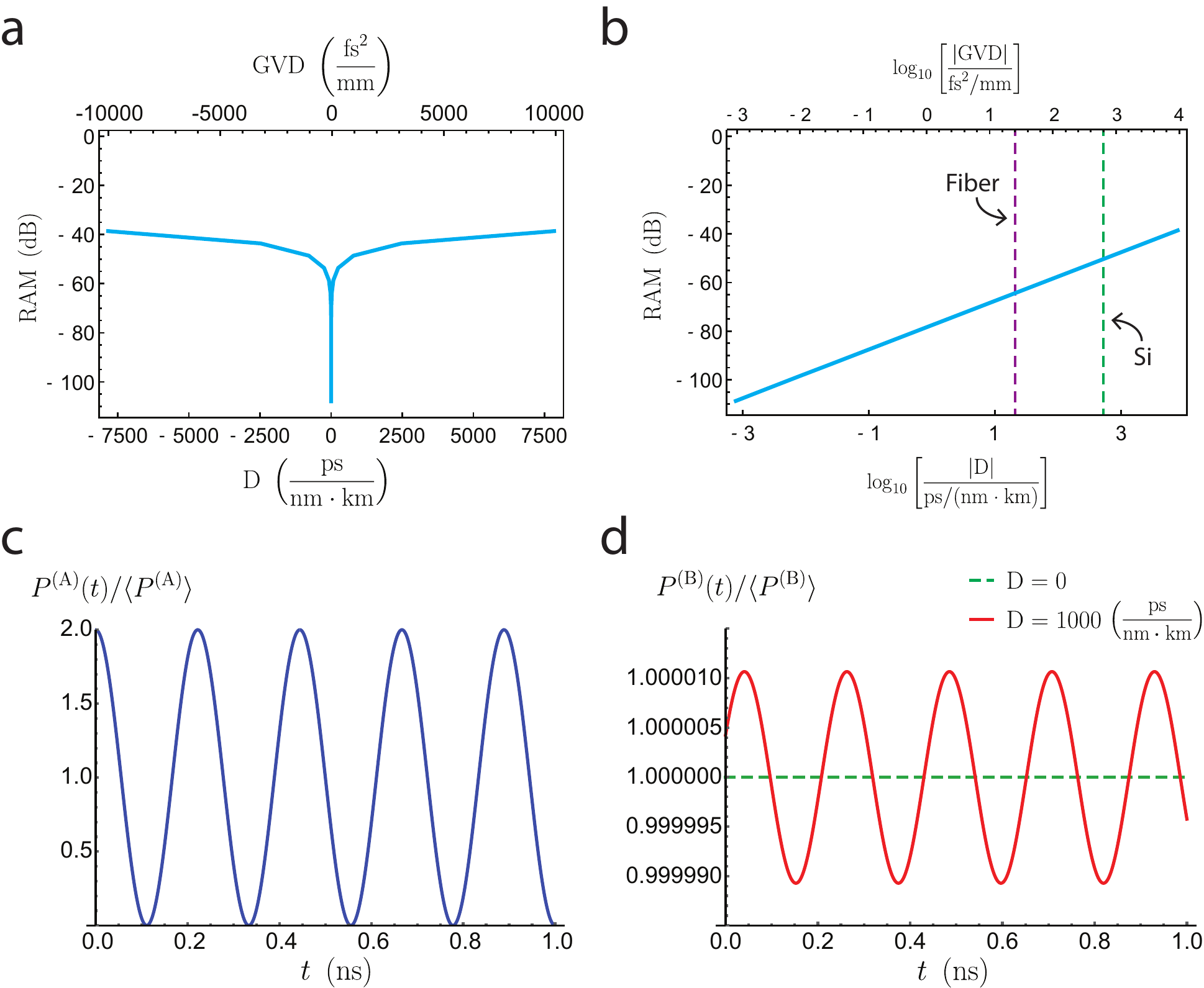}
    \caption{
    a. Numerically calculated residual amplitude modulation as a function of dispersion, for a 2.5 cm device with \(G_\text{B} = 700\text{ (Wm)}^{-1}\) and an acoustic \textit{Q}-factor of 1000. The input powers are 100 mW in each of the two tones into waveguide A and 100 mW into a single tone in waveguide B. 
    b. The same results shown on a logarithmic scale. Values of GVD in bulk silicon and optical fiber are shown for reference. 
    c. Normalized optical field envelope in waveguide A as a function of time shows the intensity modulation that drives an acoustic field at 4.5 GHz. 
    d. The normalized optical field envelope in waveguide B shows a constant envelope in the absence of dispersion (green). When some dispersion is taken into account, a small intensity modulation can be seen (red).}
    \label{fig:RAM}
\end{figure}

The numerically simulated RAM is shown in Fig. \ref{fig:RAM}, where the dispersion is given both in terms of GVD, and a commonly used dispersion parameter D, defined by 
\begin{equation}
    \begin{split}
        \text{GVD} = \frac{\partial^2 k}{\partial \omega^2} \bigg|_{\omega=\omega_0},
    \end{split}
    \qquad
    \begin{split}
        \text{D} = -\frac{2 \pi c}{\lambda^2}\text{GVD}.
    \end{split}
\end{equation}

When using numerical values consistent with practical  materials and waveguide design, the simulation results in a residual intensity modulation on the order of $-50$ dB. In many applications, the frequency response of other optical elements in the system, such as detectors, couplers and modulators, can yield similar or more dominant effects on the signal.

\section{\label{subsec:app_Classical}Classical mechanics derivation}
We can derive the equations of motion describing the dynamics of a forward-Brillouin process using a classical mechanics approach as an alternative to the derivation given earlier in Appendix \ref{subsec:app_Hamiltonian}, as summarized in the following section. We start by considering an acoustic eigen-mode, $\textbf{U}(\textbf{r})e^{-i\Omega_0 t}$ at frequency \(\Omega_0\), using \cite{royer2000elastic}
\begin{equation}
    \frac{\partial}{\partial x_j} c_{ijkl} \frac{\partial U_l}{\partial x_k} = -\rho \Omega_0^2 U_i, 
\end{equation}
where $x_i$ is a spatial coordinate, $\rho$ the mass density of the waveguide material, $c_{ijkl}$ the elastic tensor, and summation is implied for repeated indexes. The guided acoustic eigen-modes relevant to FSBS are transversely polarized, such that the acoustic mode profile can be written as $\textbf{U}(\textbf{r})=\left( U_{x}(\textbf{r}_\perp) \hat{x} + U_{y}(\textbf{r}_\perp) \hat{y}\right) e^{i q z},$   where $q$ is the acoustic wave-vector. We can write the displacement field in this waveguide as 
\begin{equation}
    \textbf{u}(\textbf{r},t) = \frac{1}{2} \left ( \sqrt{\frac{2}{\Omega_0}} \textbf{U}(\textbf{r}_\perp) b e^{i(qz-\Omega_0 t)}+ \text{c.c.} \right ),
    \label{eq:phonon_field_classical}
\end{equation}
where $b$ is the mode amplitude of the acoustic eigen-mode.

The total energy density, $\rho \dot{\textbf{u}}^2,$ associated with this displacment field is the sum of the potential and kinetic energy terms. Intergrating the total energy density over volume and making the rotating wave approximation, we get the following Hamiltonian
\begin{equation}
    \left < H_\text{0} \right > =  \Omega_0 \int{d^3x \ \textbf{U}^*(\textbf{r}_\perp) \rho(\textbf{r}_\perp) \textbf{U}(\textbf{r}_\perp) b^* b}. 
\end{equation}

We normalize the acoustic eigen-mode profile such that $\int{d^3x \ \textbf{U}^*(\textbf{r}_\perp) \rho(\textbf{r}_\perp) \textbf{U}(\textbf{r}_\perp)}=1,$ leaving us with a simplified expression for the time-averaged Hamiltonian,
\begin{equation}
    \left < H_\text{0} \right > =  \Omega_0 b^* b.
\end{equation}

Photoelastic coupling of the light and sound fields result in an interaction term in the Hamiltonian. To calculate this term, we look at the change in the energy density $\delta U=\sigma_{mn} \times  \delta S_{mn}$. Here, $\sigma_{mn}$ is the stress and $\delta S_{mn}=1/2(\partial u_m/\partial x_n + \partial u_n/\partial x_m) $ is the strain. For cubic and isotropic media, the electrostrictive stress is given by
\begin{equation}
    \sigma_{mn} = -\frac{1}{2} \epsilon_o \epsilon_r^2 p_{ikmn} E_k E_i,
\end{equation}
where $\epsilon_r$ is the relative permittivity, $p_{ikmn}$ is the photelastic tensor, and $\textbf{E}$ is the total electric field inside the waveguide. The resulting interaction Hamiltonian from the change in energy density is given by
\begin{equation}
    H_\text{I} = \int{d^3x \ \delta U} = -\frac{1}{2} \epsilon_o \int{d^3x (\epsilon_r^2 p_{ikmn} E_k E_i \delta S_{mn})},
\end{equation}
and when we consider electric fields co-polarized in the $x$-direction, this yields 
\begin{equation}
    H_\text{I} = -\frac{1}{2} \epsilon _o \int{d^3x \ \epsilon_r^2 \left( p_{11} \frac{\partial u_x}{\partial x}+p_{12} \frac{\partial u_y}{\partial y}+p_{12} \frac{\partial u_z}{\partial z}\right) E_x E_x.}
    \label{eq:int_H_class}
\end{equation}

We describe an electric field propagating in the waveguide 
\begin{equation}
    \textbf{E}_n(\textbf{r},t) = \frac{1}{2} \sum_n \left ( E(\textbf{r}_\perp) a_n e^{i(k_n z -\omega_n t)} + \text{c.c.}\right)\hat{x},
    \label{eq:electric_field_classical}
\end{equation}
where $ E(\textbf{r}_\perp)$ is the transverse optical mode profile, $a_n$ the optical mode, $k_n$ the wave vector and $\omega_n$ the angular frequency of the different optical frequencies, spaced by $\Omega$, (i.e. $\Omega = \omega_n - \omega_{n-1}$). We further assume a phase-matched process, such that $q= k_{n}-k_{n-1}$.

Substituting $\textbf{E}(\textbf{r},t)$ and $\textbf{u}(\textbf{r},t)$ into the interaction Hamiltonian from Eq. (\ref{eq:int_H_class}) we have 
\begin{equation}
    \begin{split}
        H_\text{I} &= - \frac{1}{16}\epsilon_o \sqrt{\frac{2}{\Omega_0}} \int{d^3x \ \epsilon_r^2 \left( p_{11} \frac{\partial U_{x}}{\partial x}e^{iqz}b + p_{12} \frac{\partial U_{y}}{\partial y}e^{iqz}b+\text{c.c.} \right)} \left (\sum_n E(\textbf{r}_\perp) a_n e^{i(k_n z -\omega_n t)}+\text{c.c.}\right)^2.
    \end{split}
\end{equation}

Making the rotating wave approximation, where we only consider the terms in $H_\text{I}$ that vary with frequency $\Omega$ (i.e. terms close to $\Omega_0$ that can drive the acoustic eigen-mode) we get
\begin{equation}
    \left < H_\text{I}  \right > = -\frac{1}{8} \epsilon_o \sqrt{\frac{2}{\Omega_0}} \int{d^3x \ \epsilon_r^2 \sum_n S^*(\textbf{r}_\perp)E_{n-1}^*(\textbf{r}_\perp)E_n(\textbf{r}_\perp)b^*a_{n-1}^* a_n e^{-i\Omega t} +\text{c.c.}},
\end{equation}
where the strain profile is
\begin{equation}
    S(\textbf{r}_\perp) = p_{11} \frac{\partial U_{x}}{\partial x} + p_{12} \frac{\partial U_{y}}{\partial y}.
\end{equation}

Carrying out the integration in the axial direction over the waveguide length $L$, and defining the following coupling term
\begin{align}
    g= \int{dxdy \ \epsilon_r^2(\textbf{r}_\perp) E_n^*(\textbf{r}_\perp)E_{n-1}(\textbf{r}_\perp)S(\textbf{r}_\perp)},
\end{align}
the interaction Hamiltonian resulting from the photoelastic coupling of light and sound is given by
\begin{equation}
    \left < H_\text{I}  \right > = -\frac{1}{8} \epsilon_o L \sqrt{\frac{2}{\Omega_0}} \sum_n g^* b^* a_{n-1}^* a_n e^{-i\Omega t}+\text{c.c.} 
\end{equation}

The total time-averaged Hamiltonian of the system, including the acoustic and interaction terms, is
\begin{align}
    \begin{split}
    \left < H  \right > &= \left < H_\text{0}  \right >+\left < H_\text{I}  \right > \\
     &=  \Omega_0 b^* b - \gamma \sum_n \left( g^* a_{n-1}^* a_n b^* e^{-i\Omega t}+g a_n^* a_{n-1} be^{i\Omega t}\right ),
    \end{split}
\end{align}
where we have defined $\gamma = \frac{1}{8} \epsilon_o L\sqrt{2/\Omega_0}$.

Now that we have derived all terms of the Hamiltonian of the system, we can calculate the time evolution of the acoustic field using \cite{strocchi1966complex}
\begin{equation}
    \dot{b}(t) = i\{H,b\},
\end{equation}
where the Poisson bracket is defined $\{H,b\} = ((\partial H/\partial b)(\partial b/\partial b^*)-(\partial H/\partial b^*)(\partial b/\partial b))$, yielding
\begin{equation}
    \dot{b}(t) = -i \frac{\partial H}{\partial b^*} = -i\Omega_0 b+i \gamma \sum_n g^* a_{n-1}^* a_n e^{-i\Omega t}.
\end{equation}

We now look at the envelope of the field, $b(t) = \bar{b}(t)e^{-i\Omega t}$, leaving us with the following equation of motion
\begin{equation}
    \dot{\bar{b}}(t) =i(\Omega -\Omega_0)\bar{b}+i\gamma\sum_n g^* a_{n-1}^* a_n.
\end{equation}

We add phonon dissipation by taking $\Omega_0 \rightarrow \Omega_0 - i\Gamma/2$, and see the steady state mode amplitude, i.e. $\dot{\bar{b}}(t)=0$, is given by
\begin{equation}
    \bar{b} = - \gamma\frac{\sum_n g^* a_{n-1}^* a_n}{\left( \Omega-\Omega_0+i\frac{\Gamma}{2} \right)}.
\end{equation}

Plugging back into Eq. (\ref{eq:phonon_field_classical}), the steady state displacement field associated with this phonon mode is
\begin{equation}
    \textbf{u}(\textbf{r},t) = \frac{1}{2} \left ( -\frac{\epsilon_o L}{4\Omega_0} \textbf{U}(\textbf{r}_\perp)  \frac{\sum_n g^* a_{n-1}^* a_n}{\left( \Omega-\Omega_0+i\frac{\Gamma}{2} \right)} e^{i(qz-\Omega t)} +\text{c.c.} \right ).
\end{equation}

A nonlinear polarization generated by this displacement field can scatter photons from pump to Stokes or anti-Stokes. The spatial and temporal evolution of the optical field in a medium with nonlinear polarization is described by \cite{boyd_book} 
\begin{equation}
    \frac{\partial^2E}{\partial z^2}-\frac{1}{v_p^2} \frac{\partial^2E}{\partial t^2} = \frac{1}{\epsilon_o c^2} \frac{\partial^2P^\text{NL}}{\partial t^2},
    \label{eq:P_NL}
\end{equation}
where $v_p$ is the phase velocity of the light in the waveguide, $c$ is the velocity of light in vacuum and $P^\text{NL}$ is the nonlinear polarization. Since all the electric fields are polarized along $\hat{x}$, $P^\text{NL} = \epsilon_o \epsilon_r^2 p_{11mn} \delta S_{mn} E_x$ reduces to
\begin{equation}
    \begin{split}
        P^\text{NL} &=  \epsilon_o \epsilon_r^2 \left( p_{11} \frac{\partial u_x}{\partial x} + p_{12} \frac{\partial u_y}{\partial y}+p_{12} \frac{\partial u_z}{\partial z}\right) E_x \\
        &= - \frac{ \epsilon_o^2 \epsilon_r^2L}{16\Omega_m} \left ( S(\textbf{r}_\perp)  \frac{\sum_n g^* a_{n-1}^* a_n}{\left( \Omega-\Omega_m+i\frac{\Gamma_m}{2} \right)} e^{i(qz-\Omega t)} +\text{c.c.} \right ) \left (\sum_n E(\textbf{r}_\perp) a_n e^{i(k_n z -\omega_n t)}+\text{c.c.}\right), 
    \end{split}
\end{equation}
after substituting the displacement field. Considering only the terms of nonlinear polarization that are phase matched to drive the Stokes field we get
\begin{equation}
    \begin{split}
        P_{n}^\text{NL} = -\frac{ \epsilon_o^2 \epsilon_r^2L}{16\Omega_0} \Bigg( E^*(\textbf{r}_\perp) S(\textbf{r}_\perp) & \frac{\left(\sum_n' g^* a_{n'}^* a_{n'+1} \right )}{\left( \Omega-\Omega_0+i\frac{\Gamma}{2} \right)}a_{n-1} \\
        & + E(\textbf{r}_\perp) S^*(\textbf{r}_\perp) \frac{\left(\sum_n' g a_{n'} a^*_{n'+1}\right) }{\left( \Omega-\Omega_0-i\frac{\Gamma}{2} \right)}a_{n+1} \Bigg) e^{i(k_{n} z-\omega_{n} t)} +\text{c.c.}.
    \end{split}
\end{equation}

Substituting $E_n(\textbf{r},t)$ and $P_{n}^\text{NL} $  into Eq. (\ref{eq:P_NL}) and making the slowly varying envelope approximation we get
\begin{equation}
    \begin{split}
        E(\textbf{r}_\perp) \left ( 2ik_n \frac{\partial a_n}{\partial z} + \frac{2i\omega_n}{v_{p}^2} \frac{\partial a_n}{\partial t} \right ) = \frac{ \epsilon_o^2 \epsilon_r^2 L \omega_n^2}{8\Omega_0 \epsilon_o c^2} \Bigg( E^*(\textbf{r}_\perp) S(\textbf{r}_\perp) & \frac{\left(\sum_n' g^* a_{n'}^* a_{n'+1} \right )}{\left( \Omega-\Omega_0+i\frac{\Gamma}{2} \right)} a_{n-1} \\
        & + E(\textbf{r}_\perp) S^*(\textbf{r}_\perp) \frac{\left(\sum_n' g a_{n'} a^*_{n'+1} \right )}{\left( \Omega-\Omega_0-i\frac{\Gamma}{2}  \right)} a_{n+1} \Bigg).
    \end{split}
\end{equation}

We look at the steady state (i.e. $\partial a_n/\partial t=0$) and integrate the transverse dimension after multiplying both sides by $E^*(\textbf{r}_\perp)$ to get
 \begin{equation}
    \frac{\partial a_{n}}{\partial z} = -i \alpha |g|^2 \left ( \frac{ \sum_{n'} a_{n'}^* a_{n+1'}}{\left( \Omega-\Omega_0+i\frac{\Gamma}{2} \right)} a_{n-1} + \frac{\sum_{n'} a_{n'} a^*_{n'+1}}{\left( \Omega-\Omega_0-i\frac{\Gamma}{2} \right)} a_{n+1}\right),
\end{equation}
where $\alpha = \epsilon_o \omega_s^2/\left(16k_s\Omega_0 c^2 \left < U, \rho U \right > \left <E_{n}, E_{n} \right >\right)$, and $\left < E_n, E_n\right >= \int{dx dy \  E^*(\textbf{r}_\perp) E(\textbf{r}_\perp)}.$ This result is equivalent to Eq. (\ref{eq:b}, \ref{eq:a_app}) which were derived in a quantum operator Hamiltonian approach using the Heisenberg equations of motion.

\vspace{5 mm}

In order to describe a device with multple waveguides and multiple acoustic modes, we can carry out a similar derivation, where each of the separate optical fields can be described by Eq. (\ref{eq:electric_field_classical}) in each of the waveguides. In the case of multiple acoustic modes, the acoustic field in Eq. (\ref{eq:phonon_field_classical}) can be described as the sum of the acoustic eigen-modes
\begin{equation}
\textbf{u}(\textbf{r},t) = \frac{1}{2} \sum_{m=1}^N{\left ( \sqrt{\frac{2}{\Omega_m}} \textbf{U}_m(\textbf{r}_\perp) b_m e^{i(qz-\Omega_mt)}+ \text{c.c.} \right)},
\end{equation}
where $\textbf{U}_m(\textbf{r}_\perp)$ is the \(m^\text{th}\) transverse acoustic eigen-mode, and $q$ is the acoustic wavevector in the axial direction.

\bibliographystyle{unsrtnat}
\bibliography{PPER_MAIN}

\begin{thebibliography}{39}
\providecommand{\natexlab}[1]{#1}
\providecommand{\url}[1]{\texttt{#1}}
\expandafter\ifx\csname urlstyle\endcsname\relax
  \providecommand{\doi}[1]{doi: #1}\else
  \providecommand{\doi}{doi: \begingroup \urlstyle{rm}\Url}\fi

\bibitem[Boyd(2003)]{boyd_book}
Robert~W Boyd.
\newblock \emph{Nonlinear optics}.
\newblock Academic Press, 2003.

\bibitem[Dabby and Whinnery(1968)]{dabby1968thermal}
FW~Dabby and JR~Whinnery.
\newblock Thermal self-focusing of laser beams in lead glasses.
\newblock \emph{Appl. Phys. Lett.}, 13\penalty0 (8):\penalty0 284--286, 1968.

\bibitem[Horowitz et~al.(1992)Horowitz, Daisy, Werner, and
  Fischer]{horowitz1992large}
Moshe Horowitz, Ron Daisy, Ofer Werner, and Baruch Fischer.
\newblock Large thermal nonlinearities and spatial self-phase modulation in
  {Sr$_x$Ba$_{1-x}$Nb$_2$O$_ 6$ and BaTiO$_3$} crystals.
\newblock \emph{Opt. Lett.}, 17\penalty0 (7):\penalty0 475--477, 1992.

\bibitem[Rotschild et~al.(2006)Rotschild, Alfassi, Cohen, and
  Segev]{rotschild2006long}
Carmel Rotschild, Barak Alfassi, Oren Cohen, and Mordechai Segev.
\newblock Long-range interactions between optical solitons.
\newblock \emph{Nat. Phys.}, 2\penalty0 (11):\penalty0 769, 2006.

\bibitem[Izdebskaya et~al.(2018)Izdebskaya, Shvedov, Jung, and
  Krolikowski]{izdebskaya2018stable}
Yana~V Izdebskaya, Vladlen~G Shvedov, Pawel~S Jung, and Wieslaw Krolikowski.
\newblock Stable vortex soliton in nonlocal media with orientational
  nonlinearity.
\newblock \emph{Opt. Lett.}, 43\penalty0 (1):\penalty0 66--69, 2018.

\bibitem[Shahmoon et~al.(2016)Shahmoon, Gri{\v{s}}ins, Stimming, Mazets, and
  Kurizki]{shahmoon2016highly}
Ephraim Shahmoon, Pjotrs Gri{\v{s}}ins, Hans~Peter Stimming, Igor Mazets, and
  Gershon Kurizki.
\newblock Highly nonlocal optical nonlinearities in atoms trapped near a
  waveguide.
\newblock \emph{Optica}, 3\penalty0 (7):\penalty0 725--733, 2016.

\bibitem[Sevin{\c{c}}li et~al.(2011)Sevin{\c{c}}li, Henkel, Ates, and
  Pohl]{sevinccli2011nonlocal}
S~Sevin{\c{c}}li, N~Henkel, C~Ates, and T~Pohl.
\newblock Nonlocal nonlinear optics in cold {Rydberg} gases.
\newblock \emph{Phys. Rev. Lett.}, 107\penalty0 (15):\penalty0 153001, 2011.

\bibitem[Pollard et~al.(2009)Pollard, Murphy, Hendren, Evans, Atkinson, Wurtz,
  Zayats, and Podolskiy]{pollard2009optical}
RJ~Pollard, A~Murphy, WR~Hendren, PR~Evans, R~Atkinson, GA~Wurtz, AV~Zayats,
  and Viktor~A Podolskiy.
\newblock Optical nonlocalities and additional waves in epsilon-near-zero
  metamaterials.
\newblock \emph{Phys. Rev. Lett.}, 102\penalty0 (12):\penalty0 127405, 2009.

\bibitem[Krasavin et~al.(2016)Krasavin, Ginzburg, Wurtz, and
  Zayats]{krasavin2016nonlocality}
AV~Krasavin, P~Ginzburg, GA~Wurtz, and AV~Zayats.
\newblock Nonlocality-driven supercontinuum white light generation in plasmonic
  nanostructures.
\newblock \emph{Nat. Commun.}, 7:\penalty0 11497, 2016.

\bibitem[Fakhri et~al.(2015)Fakhri, Vaziri, Jaleh, and
  Shabestari]{fakhri2015nonlocal}
P~Fakhri, MR~Rashidian Vaziri, B~Jaleh, and N~Partovi Shabestari.
\newblock Nonlocal nonlinear optical response of graphene oxide-{Au}
  nanoparticles dispersed in different solvents.
\newblock \emph{J. Opt.}, 18\penalty0 (1):\penalty0 015502, 2015.

\bibitem[Renninger et~al.(2018)Renninger, Kharel, Behunin, and
  Rakich]{renninger2018bulk}
WH~Renninger, P~Kharel, RO~Behunin, and PT~Rakich.
\newblock Bulk crystalline optomechanics.
\newblock \emph{Nat. Phys.}, 14\penalty0 (6):\penalty0 601, 2018.

\bibitem[Shin et~al.(2013)Shin, Qiu, Jarecki, Cox, Olsson~III, Starbuck, Wang,
  and Rakich]{shin2013tailorable}
Heedeuk Shin, Wenjun Qiu, Robert Jarecki, Jonathan~A Cox, Roy~H Olsson~III,
  Andrew Starbuck, Zheng Wang, and Peter~T Rakich.
\newblock Tailorable stimulated {Brillouin} scattering in nanoscale silicon
  waveguides.
\newblock \emph{Nat. Commun.}, 4:\penalty0 1944, 2013.

\bibitem[Kittlaus et~al.(2016)Kittlaus, Shin, and Rakich]{kittlaus2016_FSBS}
Eric~A Kittlaus, Heedeuk Shin, and Peter~T Rakich.
\newblock Large {Brillouin} amplification in silicon.
\newblock \emph{Nat. Photonics}, 10\penalty0 (7):\penalty0 463--467, 2016.

\bibitem[Wiederhecker et~al.(2019)Wiederhecker, Dainese, and
  Mayer~Alegre]{wiederhecker2019brillouin}
Gustavo~S Wiederhecker, Paulo Dainese, and Thiago~P Mayer~Alegre.
\newblock {Brillouin} optomechanics in nanophotonic structures.
\newblock \emph{APL Photonics}, 4\penalty0 (7):\penalty0 071101, 2019.

\bibitem[Eggleton et~al.(2019)Eggleton, Poulton, Rakich, Steel, and
  Bahl]{eggleton2019brillouin}
Benjamin~J Eggleton, Christopher~G Poulton, Peter~T Rakich, Michael~J Steel,
  and Gaurav Bahl.
\newblock {Brillouin} integrated photonics.
\newblock \emph{Nat. Photonics}, pages 1--14, 2019.

\bibitem[Van~Laer et~al.(2015)Van~Laer, Kuyken, Van~Thourhout, and
  Baets]{van2015interaction}
Rapha{\"e}l Van~Laer, Bart Kuyken, Dries Van~Thourhout, and Roel Baets.
\newblock Interaction between light and highly confined hypersound in a silicon
  photonic nanowire.
\newblock \emph{Nat. Photonics}, 9\penalty0 (3):\penalty0 199, 2015.

\bibitem[Shin et~al.(2015)Shin, Cox, Jarecki, Starbuck, Wang, and
  Rakich]{shin2015control}
Heedeuk Shin, Jonathan~A Cox, Robert Jarecki, Andrew Starbuck, Zheng Wang, and
  Peter~T Rakich.
\newblock Control of coherent information via on-chip photonic--phononic
  emitter--receivers.
\newblock \emph{Nat. Commun.}, 6:\penalty0 6427, 2015.

\bibitem[Kittlaus et~al.(2018{\natexlab{a}})Kittlaus, Kharel, Otterstrom, Wang,
  and Rakich]{kittlaus2018rf}
Eric~A Kittlaus, Prashanta Kharel, Nils~T Otterstrom, Zheng Wang, and Peter~T
  Rakich.
\newblock {RF}-photonic filters via on-chip photonic--phononic emit--receive
  operations.
\newblock \emph{J. Light. Technol.}, 36\penalty0 (13):\penalty0 2803--2809,
  2018{\natexlab{a}}.

\bibitem[Diamandi et~al.(2017)Diamandi, London, and Zadok]{diamandi2017opto}
H~Hagai Diamandi, Yosef London, and Avi Zadok.
\newblock Opto-mechanical inter-core cross-talk in multi-core fibers.
\newblock \emph{Optica}, 4\penalty0 (3):\penalty0 289--297, 2017.

\bibitem[Diamandi et~al.(2018)Diamandi, London, Bashan, Bergman, and
  Zadok]{diamandi2018highly}
H~Hagai Diamandi, Yosef London, Gil Bashan, Arik Bergman, and Avi Zadok.
\newblock Highly-coherent stimulated phonon oscillations in a multi-core
  optical fiber.
\newblock \emph{Sci. Rep.}, 8\penalty0 (1):\penalty0 9514, 2018.

\bibitem[Kittlaus et~al.(2018{\natexlab{b}})Kittlaus, Otterstrom, Kharel,
  Gertler, and Rakich]{kittlaus2018non}
Eric~A Kittlaus, Nils~T Otterstrom, Prashanta Kharel, Shai Gertler, and Peter~T
  Rakich.
\newblock Non-reciprocal interband {Brillouin} modulation.
\newblock \emph{Nat. Photonics}, page~1, 2018{\natexlab{b}}.

\bibitem[Rakich et~al.(2012)Rakich, Reinke, Camacho, Davids, and
  Wang]{rakich2012giant}
Peter~T Rakich, Charles Reinke, Ryan Camacho, Paul Davids, and Zheng Wang.
\newblock Giant enhancement of stimulated {Brillouin} scattering in the
  subwavelength limit.
\newblock \emph{Phys. Rev. X}, 2\penalty0 (1):\penalty0 011008, 2012.

\bibitem[Sipe and Steel(2016)]{sipe2016hamiltonian}
JE~Sipe and MJ~Steel.
\newblock A {Hamiltonian} treatment of stimulated {Brillouin} scattering in
  nanoscale integrated waveguides.
\newblock \emph{New J. Phys.}, 18\penalty0 (4):\penalty0 045004, 2016.

\bibitem[Rakich et~al.(2010)Rakich, Davids, and Wang]{rakich2010tailoring}
Peter~T Rakich, Paul Davids, and Zheng Wang.
\newblock Tailoring optical forces in waveguides through radiation pressure and
  electrostrictive forces.
\newblock \emph{Opt. Express}, 18\penalty0 (14):\penalty0 14439--14453, 2010.

\bibitem[Kang et~al.(2009)Kang, Nazarkin, Brenn, and Russell]{kang2009tightly}
Myeong~Soo Kang, A~Nazarkin, A~Brenn, and P~St~J Russell.
\newblock Tightly trapped acoustic phonons in photonic crystal fibres as highly
  nonlinear artificial {Raman} oscillators.
\newblock \emph{Nat. Phys.}, 5\penalty0 (4):\penalty0 276, 2009.

\bibitem[Qiu et~al.(2013)Qiu, Rakich, Shin, Dong, Solja{\v{c}}i{\'c}, and
  Wang]{qiu2013stimulated}
Wenjun Qiu, Peter~T Rakich, Heedeuk Shin, Hui Dong, Marin Solja{\v{c}}i{\'c},
  and Zheng Wang.
\newblock Stimulated {Brillouin} scattering in nanoscale silicon step-index
  waveguides: a general framework of selection rules and calculating {SBS}
  gain.
\newblock \emph{Opt. Express}, 21\penalty0 (25):\penalty0 31402--31419, 2013.

\bibitem[Kharel et~al.(2016)Kharel, Behunin, Renninger, and
  Rakich]{kharel2016_Hamiltonian}
Prashanta Kharel, RO~Behunin, WH~Renninger, and PT~Rakich.
\newblock Noise and dynamics in forward {Brillouin} interactions.
\newblock \emph{Phys. Rev. A}, 93\penalty0 (6):\penalty0 063806, 2016.

\bibitem[Wolff et~al.(2017)Wolff, Stiller, Eggleton, Steel, and
  Poulton]{wolff2017cascaded}
Christian Wolff, Birgit Stiller, Benjamin~J Eggleton, Michael~J Steel, and
  Christopher~G Poulton.
\newblock Cascaded forward {Brillouin} scattering to all {Stokes} orders.
\newblock \emph{New J. Phys.}, 19\penalty0 (2):\penalty0 023021, 2017.

\bibitem[Kang et~al.(2010)Kang, Brenn, and Russell]{kang2010all}
Myeong~Soo Kang, A~Brenn, and P~St~J Russell.
\newblock All-optical control of gigahertz acoustic resonances by forward
  stimulated interpolarization scattering in a photonic crystal fiber.
\newblock \emph{Phys. Rev. Lett.}, 105\penalty0 (15):\penalty0 153901, 2010.

\bibitem[Kittlaus et~al.(2017)Kittlaus, Otterstrom, and
  Rakich]{kittlaus2017_Intermodal}
Eric~A Kittlaus, Nils~T Otterstrom, and Peter~T Rakich.
\newblock On-chip inter-modal {Brillouin} scattering.
\newblock \emph{Nat. Commun.}, 8:\penalty0 15819, 2017.

\bibitem[Raymer and Mostowski(1981)]{raymer1981stimulated}
MG~Raymer and J~Mostowski.
\newblock Stimulated {Raman} scattering: unified treatment of spontaneous
  initiation and spatial propagation.
\newblock \emph{Phys. Rev. A}, 24\penalty0 (4):\penalty0 1980, 1981.

\bibitem[Boyd et~al.(1990)Boyd, Rza̧ewski, and Narum]{boyd1990noise}
Robert~W Boyd, Kazimierz Rza̧ewski, and Paul Narum.
\newblock Noise initiation of stimulated {Brillouin} scattering.
\newblock \emph{Phys. Rev. A}, 42\penalty0 (9):\penalty0 5514, 1990.

\bibitem[Petrov and Eich(2004)]{petrov2004zero}
A~Yu Petrov and M~Eich.
\newblock Zero dispersion at small group velocities in photonic crystal
  waveguides.
\newblock \emph{Appl. Phys. Lett.}, 85\penalty0 (21):\penalty0 4866--4868,
  2004.

\bibitem[Zhao et~al.(2012)Zhao, Zhang, Wu, and Wang]{zhao2012wideband}
Yong Zhao, Ya-Nan Zhang, Di~Wu, and Qi~Wang.
\newblock Wideband slow light with large group index and low dispersion in
  slotted photonic crystal waveguide.
\newblock \emph{J. Light. Technol.}, 30\penalty0 (17):\penalty0 2812--2817,
  2012.

\bibitem[Little et~al.(1997)Little, Chu, Haus, Foresi, and
  Laine]{little1997microring}
Brent~E Little, Sai~T Chu, Hermann~A Haus, J~Foresi, and J-P Laine.
\newblock Microring resonator channel dropping filters.
\newblock \emph{J. Light. Technol.}, 15\penalty0 (6):\penalty0 998--1005, 1997.

\bibitem[Noschese et~al.(2013)Noschese, Pasquini, and
  Reichel]{noschese2013tridiagonal}
Silvia Noschese, Lionello Pasquini, and Lothar Reichel.
\newblock Tridiagonal toeplitz matrices: properties and novel applications.
\newblock \emph{Numer. Linear Algebra Appl.}, 20\penalty0 (2):\penalty0
  302--326, 2013.

\bibitem[Otterstrom et~al.(2018)Otterstrom, Behunin, Kittlaus, Wang, and
  Rakich]{otterstrom2018silicon}
Nils~T Otterstrom, Ryan~O Behunin, Eric~A Kittlaus, Zheng Wang, and Peter~T
  Rakich.
\newblock A silicon {Brillouin} laser.
\newblock \emph{Science}, 360\penalty0 (6393):\penalty0 1113--1116, 2018.

\bibitem[Royer and Dieulesaint(2000)]{royer2000elastic}
D~Royer and E~Dieulesaint.
\newblock Elastic waves in solids, {Vol}. 1.
\newblock \emph{Springer, Berlin}, 20\penalty0 (0):\penalty0 0, 2000.

\bibitem[Strocchi(1966)]{strocchi1966complex}
Franco Strocchi.
\newblock Complex coordinates and quantum mechanics.
\newblock \emph{Rev. Mod. Phys.}, 38\penalty0 (1):\penalty0 36, 1966.

\end{thebibliography}

\end{document}